\documentclass[aps,pre,twocolumn,showpacs,superscriptaddress,reprint]{revtex4-1}

\usepackage{subfigure}
\usepackage{wrapfig}
\usepackage{textcomp}
\usepackage{inputenc}
\usepackage{enumitem}
\usepackage{graphicx}
\usepackage{amssymb}
\usepackage{amsmath}
\usepackage{bm}
\usepackage{amsmath, amsthm, amssymb}
\usepackage{hyperref}
\usepackage{color}
\usepackage{rotating}
\usepackage{afterpage}
\usepackage{tabulary,booktabs}

\usepackage{amsfonts}
\usepackage{dcolumn}
\usepackage{float}
\usepackage{bm}

\newcommand{\heavy}{\text{HZE}}

\newcommand{\darken}{\ensuremath{{\mathfrak{D}}}}
\newcommand{\udens}{\ensuremath{\,\text{g/cm}^3}}

\newcommand{\qmoy}{\ensuremath{\left < Q \right >}}
\newcommand{\Amoy}{\ensuremath{\left < A \right >}}

\newcommand{\ISO}{ {\bf \boldmath{Iso-n}\ensuremath{_e}}}
\newcommand{\iso}{ {\bf \boldmath{iso-n}\ensuremath{_e}}}
\newcommand{\dmu}{\ensuremath{\mu \text{m}^2/\text{ns}}}
\newcommand{\De}{\ensuremath{D_\text{Emp}}}
\newcommand{\Dh}{\ensuremath{D_\text{Hyd}}}
\newcommand{\Dz}{\ensuremath{D_\text{Z}}}

\begin{document}
\setcounter{table}{0}
\def\ups{Universit{\'e} Paris-Saclay, CEA, Laboratoire {M}ati{\`e}re en {C}onditions {E}xtr{\^e}mes, 91680 Bruy{\`e}res-le-Ch{\`a}tel, France}
\def\cea{CEA-DAM-DIF, F-91297 Arpajon, France}
\def\lanl{Theoretical Division, Los Alamos National Laboratory, Los Alamos, New Mexico 87545, USA}

\title{Static and dynamic properties of multi-ionic plasma mixtures}

\author{Jean Cl\'erouin}
\affiliation{\cea}
\affiliation{\ups}
%\email{jean.clerouin@cea.fr}

\author{Philippe Arnault}
\affiliation{\cea}

\author{Benoit-Joseph Gr\'ea}
\affiliation{\cea}

\author{S\'ebastien Guisset}
\affiliation{\cea}

\author{Marc Vandenboomgaerde}
\affiliation{\cea}

\author{Alexander J. White}
\affiliation{\lanl}

\author{Lee A. Collins}
\affiliation{\lanl}

\author{Joel D. Kress}
\affiliation{\lanl}

\author{Christopher Ticknor}
\affiliation{\lanl}

 \date{\today}

\begin{abstract}
Complex plasma mixtures with three or more components are often encountered  in astrophysics or in inertial confinement fusion (ICF) experiments. For mixtures containing species with large differences in atomic number $Z$, the modeling needs to consider  at the same time the kinetic theory for low-$Z$ elements combined with the theory of strongly coupled plasma for high-$Z$ elements, as well as all the intermediate situations that can appear in multi-component systems. For such cases, we study the pair distribution functions, self-diffusions, mutual diffusion and viscosity for ternary mixtures at extreme conditions. These quantities can be produced from first principles using orbital free molecular dynamics at the computational expense of very intensive simulations to reach good statistics. Utilizing the first-principles results as reference data, we assess the merit of a global analytic model for transport coefficients, "Pseudo-Ions in Jellium" (PIJ), based on an iso-electronic assumption (\iso). With a multi-component hypernetted-chain integral equation, we verify the quality of the \iso\, prescription for describing the static structure of the mixtures. This semi-analytical modeling compares well with the simulation results and allows one to consider plasma mixtures not accessible to simulations. Applications are given for the mix of materials in ICF experiments. A reduction of a multicomponent mixture to an effective binary mixture is also established in the hydrodynamic limit and compared with PIJ estimations for ICF relevant mixtures.

\end{abstract}

\maketitle

%---------------------------------------------------------------------------------------------------
\section{Introduction}
%---------------------------------------------------------------------------------------------------
Fuel-pusher mix is one of the major issues in inertial confinement fusion (ICF) experiments that  hampers the energy production goal (see e.~g. \cite{LARR18,YIN19}). Recent dedicated experiments have highlighted the importance of microscopic diffusion compared to turbulent mixing \cite{CASE14,RIND14,BAUM14,MURP16,ZYLS18}. Moreover, direct numerical simulations (DNS) of imploding targets have shown the crucial role of viscosity and diffusion  during  compression. Depending on the temperature behavior of viscosity,  a relaminarization of turbulent flow  could be expected  with possibly a massive mixing of the target by sudden diffusion \cite{WEBE14,DAVI16b,DAVI16,VICI18,VICI19}.
\begin{figure}[!h]
\begin{center}
\includegraphics[scale=0.60]{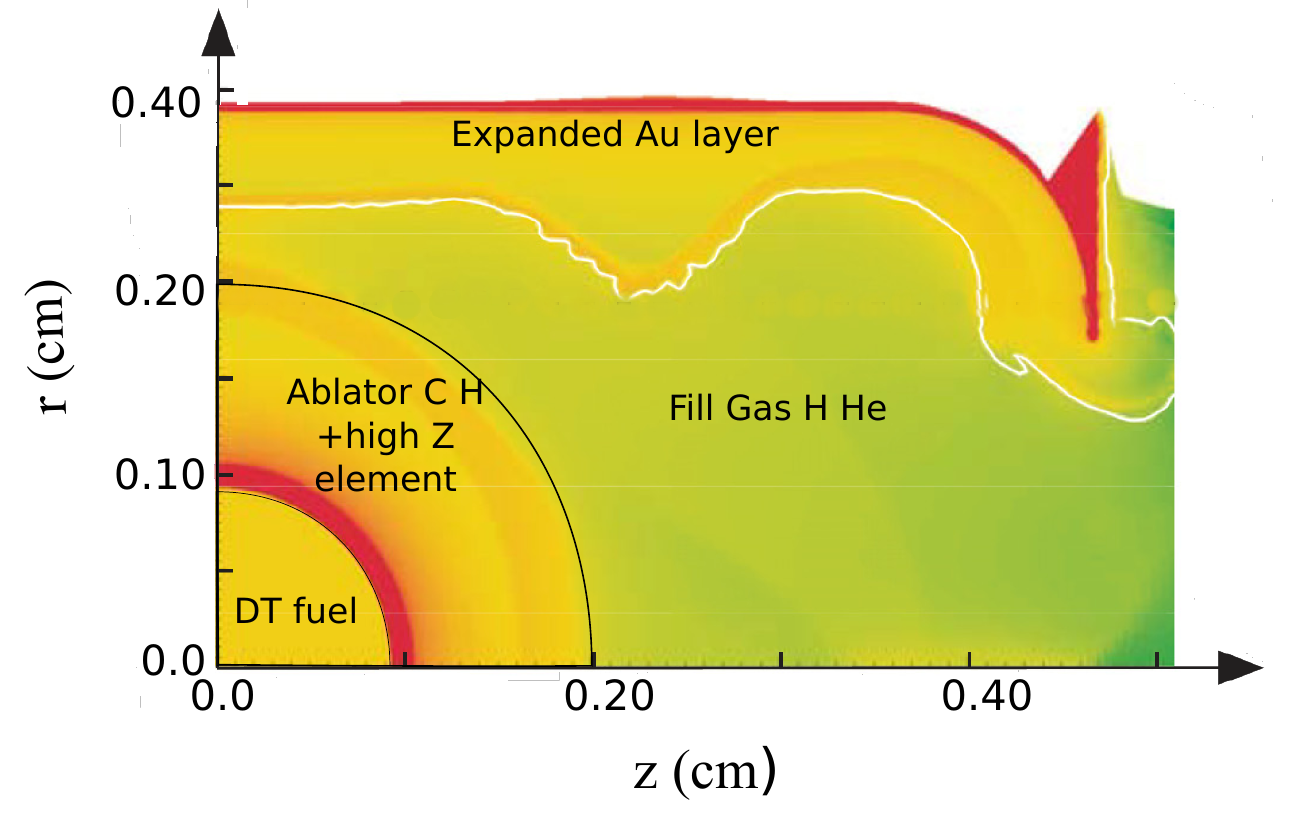}
\caption {\label{cible}(Color online) Sketch of an ICF target  showing the different elements: the DT fuel surrounded by the CH ablator, the filling gas made of He and the gold wall, during the compression process. The colour map shows the electronic density ranging from $10^{19}$ for green (light grey) to $10^{23}$ electrons/cm$^3$ for red (dark grey)  (for more details see \cite{VAND18}).}
\end{center}
\end{figure}

  An example of the different ICF mixing layers occurring in an indirect drive experiment (taken from \cite{VAND18})  is shown in Fig.\,\ref{cible}. The  capsule contains low-Z elements (LZE) deuterium (D) and tritium (T) confined within a plastic ablator  of carbon-hydrogen (CH) compounds doped with high-Z elements (HZE) germanium (Ge) or silicon or gold (Au), to improve energy couplings. In indirect drive configurations, this  capsule  is contained in a gold hohlraum often filled with a gas mixture of H/He. Under laser irradiation, the gold hohlraum expands, mixes with the gas and eventually with the ablated CH. Inside the capsule, the DT fuel is also subjected to mix with CH.

An important feature of asymmetric mixtures such as He/Au, is  that each component  experiences a very different plasma coupling and cannot be taken on the same footing. By coupling, we mean the non-negligible effect of the Coulomb interactions between the ions. A convenient measure of correlation in one-component plasma is the coupling parameter $\Gamma$ defined in atomic units by
\begin{equation}
\Gamma=\dfrac{Q^2}{a T_i}.
\end{equation}
In this expression, $Q = Q(n_e, T_e)$ is the actual charge (the ionization degree), which depends on the electron density $n_e$ and temperature $T_e$, $a=(3/4 \pi n)^{1/3}$ the ionic mean sphere radius, $n$ the total atomic density, and $T_i$ the ion temperature. The coupling parameter $\Gamma$ represents the ratio of the interparticle potential energy to the kinetic energy of the particles. A realistic description of a hot dense plasma, where $\Gamma \geq 1$, requires a model that includes all interactions explicitly (N-body problem) in contrast to weakly-coupled plasmas, where $\Gamma \ll 1$, that are well represented by binary collisions and mean-field effects. A rough estimation of the coupling parameters of the various constituents of the target is given in Table~\ref{table1}, showing the difference of the coupling parameter $\Gamma$ between CH, He, and Au at the same temperature. Helium is purely kinetic ($\Gamma\simeq 0.03$) when gold is in the strongly coupled regime ($\Gamma\simeq 5$). CH is in a moderate coupling regime. We shall discuss in more details the concept of coupling parameter in mixtures. 
\begin{table}[h]
\caption{Plasma parameters deduced from thermodynamic conditions ($\rho$, $T$) for the layers occurring in indirect drive experiments. DT fuel is not treated here. Data are taken from a hydrodynamic simulation during the compression.}
\begin{center}
\begin{tabular}{|c|c|c|c|}
			& CH		&	He			&	Au		\\
\hline\hline
$\rho$  \udens	& 5 10$^{-3}$	&	3 10$^{-3}$	&	1 10$^{-2}$\\
T  eV		& 200		& 200			& 200		\\
\qmoy		& 3.5			& 2				&38			\\
$\Gamma$	&	0.1		& 0.03			&5.3			\\

\end{tabular}
\end{center}
\label{table1}
\end{table}%

We shall discuss mixtures containing both LZE and HZE. In such asymmetric mixtures the two components behave very differently. When the LZE is quickly fully ionized and becomes more and more weakly coupled as the temperature rises, the HZE ionizes continuously, by letting more and more electrons go into the continuum and remains strongly coupled up to about ten keV. Under certain conditions, it can even exhibit a constant coupling behavior, the ``$\Gamma$-plateau'' \cite{CLER13b,ARNA13,FAUS20}, characterized by  a static structure insensitive to the temperature. These extreme behaviors warrant joining two different approaches:  kinetic theory for the LZE and a modeling of strongly coupled plasmas to describe the HZE. The combination of these two approaches is the backbone of the pseudo-ion in jellium (PIJ) model to describe transport in plasma mixtures \cite{ARNA13b}. 
%Adding a third, fourth etc. components rises the question of gathering some components into an effective average species such as CH.

As stated before, in ICF experiments,  complex mixtures with three, four, or five components are often encountered. While such complex multi-component mixtures represent a challenge for theory, they are more easily amenable to orbital-free molecular dynamics (OFMD) simulations \cite{LAMB07, LAMB13}, in which the ionization states of the various components  are not an input parameter but evolve self-consistently during the simulation.  Previous extensive OFMD simulations of binary mixtures demonstrated the influence of the  HZE on the static and dynamic properties of a LZE \cite{TICK16, WHIT17}. In particular,  the crowding of the LZE  was evidenced in OFMD simulations, and reproduced by a multicomponent hypernetted chain (MCHNC) integral equations approach. The consequences on the enhancement of nuclear reactions rates were evaluated \cite{ARNA19,CLER19}. We also showed that small amounts of HZE can substantially lower the viscosity of the mixture  and the LZE self-diffusion \cite{TICK16,WHIT17}.  Concerning mutual diffusion coefficients, we assessed the validity of the Darken approximation, connecting the mutual diffusion to the self diffusion coefficients of the species in the mixture. These simulation results on transport coefficients allowed us to tune the PIJ model \cite{ARNA13b}, over a wide range of thermodynamic conditions. We shall discuss in more details the corrections to the PIJ model involved in this tuning.

In this paper, we tackle the modeling of ternary mixtures.  We consider a mixture made of a LZE with a medium-Z element (MZE), \textit{e.g.} the CH ablator constitutes such a mixture. We investigate  the properties of the CH compound mixed with a HZE element, which can be provided by Si or Ge from the dopant or by Au  from the hohlraum. 

The question is: how the mutual diffusion or the viscosity of the H-C mixture is modified by an increasing amount of gold? Does the addition of the HZE enhance or hamper the mutual diffusion between C and H elements? 
To shed light on these issues, we shall analyze results of extensive OFMD simulations on ternary mixtures \cite{WHIT19}, retaining simulations on CHAg as a challenging test-case for the PIJ model and as  the prototype of ternary mixtures of interest to ICF. This extends the test-bed of our modeling beyond binary systems. In the first part, we shall recall the main features of the OFMD simulations, the PIJ model, and the MCHNC calculations. We shall then assess the validity of the iso-electronic assumption (\iso\,), which is used to determine ionizations in particular, first comparing the static structures of various mixtures  and finally the transport properties of mixtures simulated with OFMD in \cite{WHIT19}.  We first test the modeling  on   ternary mixtures of HXAg, at 200 and 400~eV with X=D, He, Li, C, Na, K, Cr, Cu and Rb.
Then, at given temperatures of 100, 200 and 400 eV, we show the influence of  increasing proportion of HZE, Ag, in equimolar binary mixtures of CH. We also discuss the validity of an effective binary mixture built with the introduction of an effective component with averaged characteristics to replace two species such as CH. We end up examining mixing layers occurring in ICF experiments.

A mixture is defined  by $M$ species, each  containing $N_i$ atoms. The total number of atoms is $N=\sum_{i=1}^{M} N_{i}$. Number concentrations are $x_i=N_i/N$ for each species. For convenience, we will particularize the number concentration of HZE $x=N_\heavy/N$.  Mixtures will be denoted by their concentrations in numbers  ($x_1$:~$x_2$:~$x_3$). In particular, we shall analyze the behavior of an equimolar binary mixture with a varying proportion of a third element ($x_1=x_2$ and $x_3=1-2x_1$).

  %--------------------------------------------------------------------------------------------------------------------------------------------------------------------------------
\section{OFMD simulations}
%--------------------------------------------------------------------------------------------------------------------------------------------------------------------------------
The OFMD method \cite{LAMB07,LAMB13} can bear the numerical burden of the simulation of a multi-component system at high temperature, in a reasonable amount of computer time. 

In these simulations, the charges of each species, $Q_1, Q_2$ ...., are not input data, but are  the result of the screening of the bare coulomb potential by the local polarization of the electronic density, which evolves at each time step according to the configuration of the nuclei. For this reason, OFMD belongs to the family of methods based on Density Functional Theory (DFT) in contrast to classical molecular dynamics simulations that use effective potentials. The method is described at length in Ref.\,\cite{LAMB13} and involves a Thomas-Fermi finite temperature expression for the non-interacting electron free energy, which bypasses the corresponding Kohn-Sham orbital-based expression. The exchange-correlation energy of the electrons is included using the Perdew-Zunger functional in the local density approximation \cite{PERD81}. This {\it density-only} functional theory allows for the exploration of high-temperature situations with almost no limitations. However, the simulations of mixtures are still limited by the dissimilar collision frequencies between low and high atomic number elements. In practice, we observed that it is difficult to go beyond  a few keV for a very asymmetric mixture due to a reduction of the time-step  as low as 0.1\,a.u. (2.42\,10$^{-18}$ s) to ensure good energy conservation.

 Molecular dynamics gives access to the static and dynamic correlations. The most simple correlation function is the pair distribution function (PDF), which describes the microscopic structure among the nuclei and gives the conditional probability $g(r)$ to have two particles at a distance $r$ with respect to a non-interacting system.
%It is obtained as the ratio of the number density of species $i$ to the ideal gas density at a distance $r$ away from a particle of species $j$. 
In particle simulations with different species $N_i$, $N_j$, $...$ , it is obtained by averaging histograms accumulated on several snapshots according to
\begin{equation}
%g_{ij}(r) = \dfrac{V}{N_i (N_j-\delta_{ij})}\,\left<\sum_{p=1}^{N_i}\,\mathop{{\sum}'}_{q=1}^{N_j}\,\delta(\mathbf{r}-\mathbf{r}_{pq}^{(ij)})\right>,
g_{ij}(r) = \dfrac{V}{N_i (N_j-\delta_{ij})}\,\left<\sum_{p=1}^{N_i}\,\mathop{{\sum}'}_{q=1}^{N_j}\,\delta(\mathbf{r}-(\mathbf{r}_p^i-\mathbf{r}_q^j)) \right>,
\end{equation}
where the primed sum excludes $q = p$ if $i=j$ and the angular brackets denotes an ensemble average. We shall compare these quantities resulting from OFMD with the solutions of the MCHNC integral equations using effective potentials.

The ionic transport coefficients are obtained from an integration in time of dynamic correlation functions. For instance, the Green-Kubo formalism relates the self-diffusion coefficient of a particular ion species, $D_{i}$, to the integral of the velocity autocorrelation function (VACF)
\begin{equation}
D_i =  {\frac{1}{3}} {\frac{1}{N_i}} \int_0^{\infty} dt  \langle {\sum_{{\alpha}=1}}   {{\vec{v}}_{\alpha}^{i}} (0) \cdot   {{\vec{v}}_{\alpha}^{i}}(t) \rangle.
\end{equation}
with ${{{\vec{v}}_{\alpha}}^{i}}$, the velocity of the  $\alpha$-th particle of species i.
The self-diffusion only depends on contributions from the same particle of the same species. On the other hand, the mutual diffusion coefficient and viscosity include the cross-correlation terms between different particles of the same and of different species.

The mutual diffusion coefficients D$_{ij}$ between species i and j derive from the Onsager coefficients ${\Lambda}_{ij} $ \cite{WHIT19}
\begin{equation}
%{\Lambda}_{ij}       \stackrel{t \rightarrow \infty}{\longrightarrow}        {\frac{1}{3}} {\frac{1}{N}} \int_{0}^{t} dt^{\prime}  \langle {{\bf{V}}_i}(0) \cdot {{\bf{V}}_j}(t^{\prime}) \rangle,
{\Lambda}_{ij}       =       {\frac{1}{3}} {\frac{1}{N}} \int_{0}^{\infty} dt  \langle {{\bf{V}}_i}(0) \cdot {{\bf{V}}_j}(t) \rangle,
\label{eq:VAC}
\end{equation}
which in turn depend upon the total species velocity $ {{\bf{V}_i}}(t)  = \sum_{{\alpha}=1} {{{\vec{v}}_{\alpha}}^{i}} (t)$.
A binary system yields only one mutual diffusion coefficient D$_{12}$, which is directly related to the $\Lambda_{12}$ Onsager coefficient and number fractions by
\begin{equation}
{{D}}_{12} = -\left [\left ({\frac{x_2}{x_1}} \right )^2 + \left ({\frac{x_1}{x_2}}\right )^2 + 2 \right ] \Lambda_{12} .
\end{equation}
However, for ternary and higher systems, the relationship among diffusion and Onsager coefficients becomes more complicated  \cite{WHIT19}.

Neglecting cross terms in the correlation function, we can relate the mutual diffusion to the self-diffusion coefficients. The resulting relation is known as the Darken approximation, which reads \cite{WHIT19}
\begin{eqnarray}
   \label{Darken1}
D_{ij}&=&{\frac {D_iD_j}{D_{mix}}} \\   
 {\frac{1} {D_{mix}}}&=&\sum_i^{M}{\frac {x_i}{D_i}}  .
   \label{Darken2}
\end{eqnarray} 
In this formula, the self-diffusion coefficients $D_i$ must be evaluated in the mixture. As shown in the previous study \cite{WHIT19}, this approximation is usually valid for high density and temperature regimes to within better than a factor of two and many times to within few percent, thus  furnishing a convenient framework to interpret diffusion in multicomponent systems.

The shear viscosity $\eta$ is computed from the autocorrelation
functions of the stress tensor components $P_{xy}$, $P_{xz}$, and $P_{yz}$ 
\begin{eqnarray}
\label{STACF}
\eta={1\over V k_BT}\int_0^\infty \langle P_{xy}(t) P_{xy}(0) \rangle \, dt.\\\nonumber
P_{xy}(t)=\underset{i}{\sum}~\left[m_i \dot x_i(t) \dot y_i(t)
               + ~y_{i}(t) F_{ix}(t)\right],
\end{eqnarray}
where the summations run over all the particles, and $\mathbf F_{i}$ is the force acting on particle $i$, negative of the gradient of the potential energy surface $V(\mathbf r)$. 

Whereas the determination of the diffusion coefficients involves the velocities, the viscosity needs in addition a potential contribution through the forces. In kinetic theory, there is only a velocity contribution since these forces are neglected at weak coupling. For this reason, the viscosity departs from the kinetic predictions earlier than the diffusion when the coupling increases.

We have performed OFMD simulations of mixtures H-X-Ag with X=D, He, Li, C, Na, K, Cr, Cu, and Rb  at 100 eV, 200 eV and 400~eV with number fractions x$_H$ = 0.4, x$_X$ = 0.4, and x$_{Ag}$ = 0.2. For each temperature, we maintain a constant total pressure by varying the mass density. For example, for 200\:eV, we fix the pressure at 720 Mbar by changing the density from 16.08\,g/cm$^3$ for D to 18.9\,g/cm$^3$ for Rb  (see Table\,\ref{table2}).
The simulations contained two hundred (200) atoms with trajectories of 10$^5$ time steps of length 0.012 fs. The Fast Fourier Transform grid consisted of 256$^3$ points. We have also examined the HCAg mixture in more detail by taking equal parts of H and C and varying the Ag concentration at a constant pressure of 720\,Mbar, maintained by varying the density. Similar parameters to those noted above served to guarantee convergence of the basic static and transport properties to better than 10\%.

 %--------------------------------------------------------------------------------------------------------------------------------------------------------------------------------
\section{Global semi-analytic models}
%------------------------------------------------------------------------------------------------------------------------------------------------------------------------------------
Although OFMD simulations are a few orders of magnitude faster than conventional orbital-based DFT simulations, some simulations still require large computational investments to obtain converged quantities.  Thus, developing a global model able to predict instantaneously transport coefficients becomes desirable.

To this end, we propose the PIJ model, which starts by introducing a simplified picture of plasma defined by ionizations and Coulomb couplings, 
through an iso-electronic \iso\, assumption.

%---------------------------------------------------------------------------------------------------
\subsection{The iso-electronic assumption}
%---------------------------------------------------------------------------------------------------
To set the ionizations $\{Q_i\}$ for each species $i$, we impose a constant electronic density  in the volume considered as the superposition of atomic volumes of radii $\{a_i \}$ $V_i=\sum_i N_i {4 \over 3} \pi a_i^3$.   The neutrality of each atomic sphere  of radius $a_i$ 
%\begin{equation}
%n_e= {\frac{Q_1(a_1,T)}{\frac{4}{3}\,\pi\,a_1^3}}={\frac{Q_2(a_2,T)}{\frac{4}{3}\,\pi\,a_2^3}} ={\frac{\qmoy}{\frac{4}{3}\,\pi\,a^3}} =\dots .
%\end{equation}
leads to the simple equations 
\begin{equation}
\label{iso}
{\frac {a_1}{Q_1^{1/3}}} = {\frac {a_2}{Q_2^{1/3}}} =\dots = {\frac {a}{\qmoy^{1/3}}},
\end{equation}
where $\qmoy=\sum_i x_i Q_i$ is the average charge. The ionizations $Q_i$ for each species  are evaluated at the density corresponding to the volume $a_i$ and at temperature $T$ (we consider $T_e=T_i$). Using More's fit  of the  Thomas-Fermi finite temperature average ionization \cite{MORE83,MORE85}, and starting from $a_1=a_2=\dots=a$, the solution is quickly reached in a few iterations. This \iso\, assumption is well known in the opacity community \cite{LEE18}.
\begin{figure}[!h]
\begin{center}
\includegraphics[width=4.0cm]{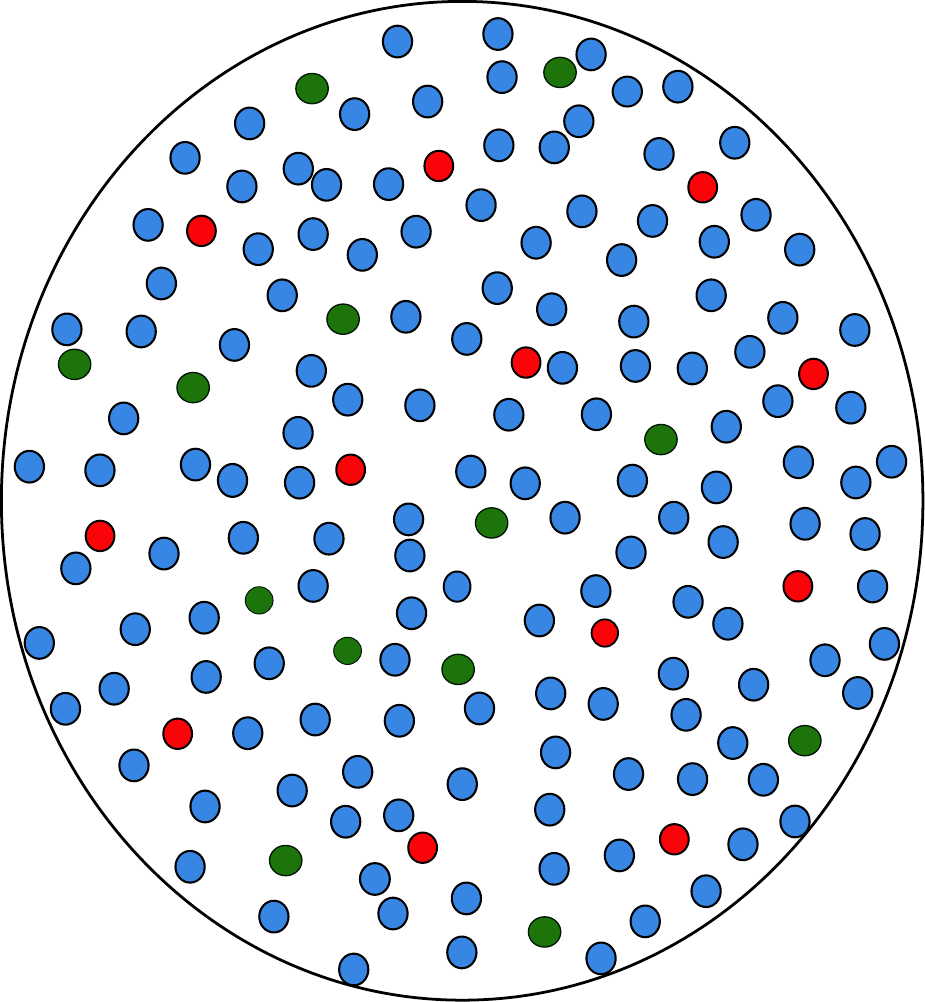}
\includegraphics[width=4.0cm]{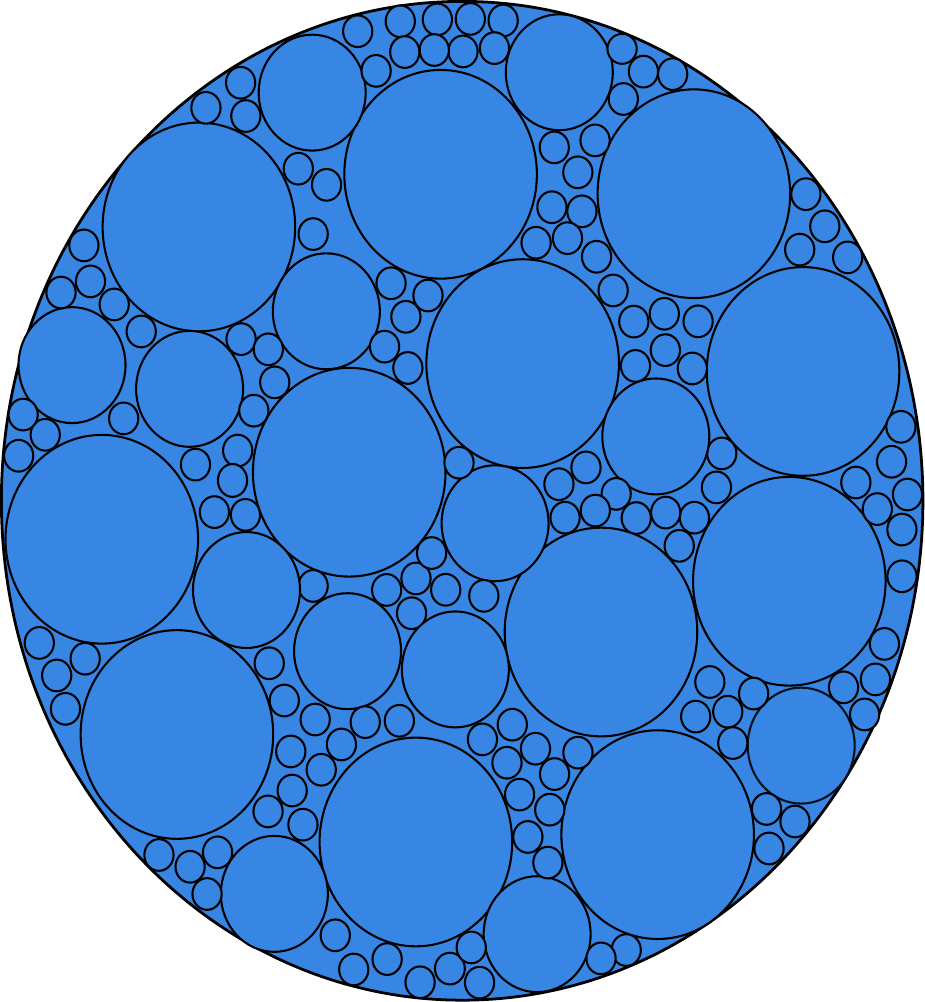}
\caption{(Color online) The \iso\, concept. Left: a non-interacting three component mixture with three different increasing atomic numbers, represented by the colors blue, green and red. Right: the radii $\{a_i\}$, and the charge states $\{Q_i\}$  are chosen to ensure a constant electronic density (light blue). The dissymmetry of radii is exaggerated for demonstration.}
\label{fig:cartoon}
\end{center}
\end{figure}

Fig.\,\ref{fig:cartoon} illustrates the \iso \, concept for a three component mixture. On the left: a non-interacting three component mixture with three different increasing atomic numbers is represented by the colors blue, green and red. On the right: the \iso\, procedure sets  the radii $\{a_i\}$ and the charge states $\{Q_i\}$  of each species to ensure the same global average value of the electronic density (light blue). Radii on the Fig.\,\ref{fig:cartoon} are exaggerated to show that by switching the charges on, the interactions then structure the system.

This concept leads to a steric representation of the system in terms of {\it big} and {\it small} ions and explains the caging effect experienced by the {\it small} ions constrained by the  {\it big} ions as illustrated by the cartoon of Fig.\,\ref{fig:cartoon}. This caging effect, which is a mixture effect,  is responsible for an extra contribution to the enhancement factor of nuclear reactions rates between low atomic number species \cite{CLER17,CLER19,ARNA19}.

In the weakly coupled regime the \iso\, prescription is a crude estimate since the neutralizing sphere is the Debye sphere containing many ions in this regime, not the Wigner-Seitz sphere of neutral pseudo-atom. A more physically-sound characterization of the distribution of charges is given by the Saha equations \cite{SALZ98}. Each species is then represented by several charge states.
%---------------------------------------------------------------------------------------------------
\subsection{Coupling parameters in a mixture}
%---------------------------------------------------------------------------------------------------
%%....................................................................................................................
\subsubsection{Effective multi-component plasma}
%%....................................................................................................................
In the same spirit we introduce the concept of effective one component plasma (EOCP) for pure elements \cite{CLER16,WANG20}, with an effective ionization inclusive of the electron screening at short range, we search for an effective multi-components model that reproduces the structure.  Indeed, a clear evidence was given that the ion-ion interaction at short range is of Coulomb type whereas a Debye-H{\"u}ckel form, with the electron screening, is recovered at long range \cite{CLER16}. The generalization of the OCP to a multicomponents system is known as binary ionic mixtures (BIM) for two components, and ternary ionic mixtures (TIM) for three components, and so on.

The previous \iso\, assumption defines a coupling parameter for each species \cite{Comment_Gam}  using the ionic radii $\{ a_i\}$ as characteristic length scales and the charges $\{Q_i\}$ (we consider here only self-couplings)
\begin{eqnarray}
\Gamma_{1}	&=&{\frac {Q_1^2}{a_1 T}} \\ 
\Gamma_{2}	&=&{\frac {Q_2^2}	{a_2 T}}\\
\Gamma_{3}	&=&{\frac {Q_3^2}	{a_3 T}}.
\end{eqnarray}
These coupling parameters are used in the PIJ model. 

We consider that the third element is the HZE with coupling 
\begin{equation}
\Gamma_\heavy={\frac {Q_\heavy^2}{a_\heavy T}},
\end{equation}
that can be rewritten  using Eq.~(\ref{iso}) as
\begin{equation}
\Gamma_\heavy	={\frac {Q_\heavy^{5/3}\qmoy^{1/3}  }	{a T}}. 
\label{GHZE}
\end{equation}
This notation shows  that the coupling parameter of the HZE comes partly from the charge itself of the considered species and partly from the surrounding medium characterized by $\qmoy$. Eq.~(\ref{GHZE}) introduces also the global quantity $a$, the Wigner-Seitz radius, which depends only on the total number of atoms.

The PIJ model also uses a global coupling parameter 
\begin{equation}
\Gamma_\text{eff}=\sum_i x_i \Gamma_i={\frac {\left < Q^{5/3} \right > \qmoy^{1/3}}{a T}}.
\end{equation}
%%....................................................................................................................
\subsubsection{Corrections to the effective OCP}
%%....................................................................................................................
In previous papers \cite{TICK16,WHIT17}, we found that PIJ self-diffusion coefficients were underestimated for HZE  at low concentration ($x \rightarrow 0$). This underestimation corresponds to an overestimation of the coupling parameter of the HZE in the dilute regime. To fix this, instead of using $Q_\heavy^{5/3}\qmoy^{1/3}$ for the definition of the coupling of HZE, we use $Q_\heavy^{2-\alpha} \qmoy^{\alpha}$ where $\alpha$ allows the screening of  the charge $Q_\heavy$ and the transfer of a part of it to the average plasma $\qmoy$. The value $\alpha=1$ provides a lower coupling $\Gamma_\text{HZE}$, which corrects the underestimated HZE self-diffusion. To obtain a smooth transition between    the limits $x \rightarrow 0$ and  $x \rightarrow 1$,   we can introduce a linear interpolation  $\alpha=-2x/3+1$  with the concentration $x=N_{\heavy}/N_\text{Tot}$. Here, we use 
\begin{equation}
\Gamma_\heavy={\frac {Q_\heavy^{2-\alpha}\qmoy^{\alpha}}{a T}}.
\label{lab:correc1}
\end{equation}

This renormalization is applied for the computation of all components of self-diffusion to account for dilute situations when the strong coupling contribution is not negligible.
%---------------------------------------------------------------------------------------------------
\subsection{Multicomponent hyper-netted chain equations}
%---------------------------------------------------------------------------------------------------
The modeling of the static structure is provided by the MCHNC procedure which  rests on the multicomponent generalization of the integral equations of the theory of fluids, namely the hyper-netted chain theory (HNC). The PDF is the central quantity in MCHNC.
For single species, the integral equations approach is based on the general Ornstein-Zernike (OZ) relation \cite{BAXT68,HANS06}
\begin{equation}
h(r)=c(r)+n \int c\left (  |\mathbf{r}-\mathbf{r}'| \right )h(\mathbf{r}') d\mathbf{r}',
\label{oz1}
\end{equation}
in which $h(r)=g(r)-1$ and $c(r)$ is the direct correlation function. The OZ relation introduces the direct correlation between two particles separated by $r$ (first term), and relates this function to the total correlation function $h(r)$ by accounting for the correlations due to all of the other pairs (second term). To determine $g(r)$ from a two-body potential $V(r)$,  a closure relation is necessary 
\begin{equation}
g(r)=\text{exp}[-\beta V(r)+ h(r)-c(r)+B(r)],
\label{oz2}
\end{equation}
where $\beta=1/kT$. $V(r)=Q^2/r$ is the Coulomb potential, which must not be confused with the self-consistent potential computed in OFMD for the forces.
 $B(r)$ is the bridge function representing three-body and higher order correlations. In the original HNC scheme, $B(r)$ is neglected. For coupling parameters less than 20, this correction is marginal and is thus not considered here.
%When needed, ($\Gamma > 10$) we added the bridge term $B(r)$ proposed by Iyetomi at al. \cite{Iyetomi1989}.

 The generalization to systems of $M$ different species leads to as many OZ relations and closures as pair interactions between species \cite{HANS06,ROGE80}
 \begin{eqnarray}
h_{ij}(r)&=&c_{ij}(r)+n \sum_{k =1}^M x_k \int c_{ik}\left (  |{\mathbf{r}}-{\mathbf{r}}'| \right )h_{kj}({\mathbf{r}}') d\mathbf{r}' \nonumber  \\
g_{ij}(r)&=&\text{exp}[-\beta V_{ij}(r)+ h_{ij}(r)-c_{ij}(r)+B_{ij}(r)],
\label{oz1m}
\end{eqnarray}
with $V_{ij}(r)=Q_iQ_j/r$ the effective potential.

The inputs of the  MCHNC calculations are the charges, concentrations, density, and temperature. 
The charges $\{Q_i\}$ are obtained through the $\iso$  prescription given by Eq.~(\ref{iso}).

%---------------------------------------------------------------------------------------------------
\subsection{PIJ Model}
%---------------------------------------------------------------------------------------------------

The PIJ model addresses the calculation of viscosity and diffusion for plasma mixtures of an arbitrarily large number of components, across Coulomb coupling regimes, and  connects  kinetic theory  with strongly- coupled plasmas models.  In the weakly-coupled regime, it is based  on a  relaxation time approximation to cope with the multi-component issue. For binary mixtures, more accurate kinetic formulas exist to account for  the distortion of the  Maxwellian distribution in the presence of concentration gradients \cite{KAGA14,STAN16,SIMA16, SHAF17}. This can multiply the mutual diffusion by  a factor, named the \emph{ relaxation correction}, varying from 1 to around 4 according to the mass ratios $m_i/m_j$ and the Coulomb coupling \cite{SHAF17}.  Unfortunately, no such accurate formulation exists for more than two components, to our knowledge.

In the strongly-coupled regime, PIJ relies on the properties of the OCP using mixing laws adapted to each transport coefficient. The connection between both regimes is performed, first, by extrapolating the kinetic formulas into the strongly-coupled regime (with a threshold of the Coulomb logarithm), and, second, by adding the corrections from the OCP quantities that arise in excess of the kinetic contribution at large coupling.

The mutual diffusion coefficients can be written
\begin{equation}
D_{i j} = D_{i j}^\text{kin} + \Delta D_{i j}^\text{ex},
\end{equation}
with \cite{BRAG65,DECO98,KAGA14}
\begin{equation}
D_{i j}^\text{kin} = 1.19 \dfrac{x_j}{m_i}~\dfrac{k_B T}{\nu_{i j}},
\end{equation}
where $\nu_{i j}$ is the collision frequency for species $i$, considering its collisions with species $j$, defined below. 
The excess correction  $\Delta D_{i j}^\text{ex}$ is computed adapting to excess quantities  the Darken relation of binary mixtures
\begin{equation}
\Delta D_{i j}^\text{ex} = x_i\, \Delta D_j^\text{ex} + x_j\, \Delta D_i^\text{ex}.
\end{equation}
 For each self-diffusion $D_i$, we use the OCP formulations with the previous coupling parameters  $\Gamma_i = Q_i^2/ a_iT$\begin{equation}
 \Delta D_i^\text{ex} = D_\text{OCP}(\Gamma_i) - D_\text{OCP}^\text{kin}(\Gamma_i).
 \end{equation}

The self-diffusion coefficients have the form
\begin{equation}
D_{i } = D_{i}^\text{kin} + \Delta D_{i}^\text{ex},
\end{equation}
with
\begin{equation}
D_{i }^\text{kin} = \dfrac{1.19}{m_i}~\dfrac{k_B T}{\nu_{i }},
\end{equation}
where $\nu_i = \sum_{j=1}^M \nu_{i j}$ is a collision frequency for species $i$, considering its collisions with all the species present in the plasma.

 The shear viscosity becomes
\begin{equation}
\eta = \eta_\text{kin} + \Delta \eta_\text{ex},
\end{equation}
with  \cite{BRAG65,DECO98}
\begin{equation}
\eta_\text{kin} =   0.965\sum_{i=1}^M \, \dfrac{n_i k_B T}{\nu_i}.
\end{equation}
The excess correction uses an equivalent OCP with an effective coupling parameter for the whole mixture $\Gamma_\text{eff} = \sum_{i=1}^M x_i \, \Gamma_i$, 
\begin{equation}
\Delta \eta_\text{ex} = \eta_\text{OCP}(\Gamma_\text{eff}) - \eta_\text{OCP}^\text{kin}(\Gamma_\text{eff}).
\end{equation}
References \cite{ARNA13b,BAST05,DALI06} contain more details on the parametrizations of the OCP properties that are used in the PIJ model.

At the lowest order of approximation, the collision frequencies are given by a Maxwellian estimate. Using the Fokker--Planck--Landau (FPL) kinetic equations  we find \cite{DECO98}
\begin{equation}
\label{nuKin}
\nu_{i j} = \dfrac{n_j}{m_i}~\dfrac{4 \sqrt{2 \pi~m_{i j}}~Q_i^2 ~ Q_j^2~e^4~\ln \Lambda_{i j}}{3~\left(k_B T\right)^{3/2}},
\end{equation}
where $\ln \Lambda_{i j}$ is the Coulomb logarithm for binary collisions between species $i$ and $j$ and $m_{i j} = m_i\,m_j/(m_i+m_j)$ is the reduced mass.
To account for the relaxation corrections in the mixture as well as for pure elements, we introduced corrections factors to the collision frequencies. These factors need only be tuned once to calibrate the PIJ model with respect to the mixtures under study.

For the studied cases of HCAg mixtures, we have obtained good results by multiplying the PIJ collision frequency between H and C by a factor of 0.4, between C and Ag by a factor 0.4 and between H and Ag by 0.2. These factors  are close to the friction coefficients $A_{lh}$ introduced by Kagan in a Chapman-Enskog calculation for the binary mixture case \cite{DECO98,KAGA14, SHAF17}. Collecting previous data for various mixtures of different mass ratio and comparing with Kagan's results (Fig.\:1 of ref \cite{KAGA14}), suggests a simple approximate relation for the friction coefficient
\begin{equation}
A_{ij}=X_{ij}^{-1/3},
\end{equation}
where $X_{ij}=A_j/A_i$ is the mass-ratio of  a couple of elements ($i,j$).
We have successfully tested this parametrization on DLiAg mixtures \cite{WHIT19} and on our previous studies of binary mixtures \cite{TICK16,WHIT17}.

%---------------------------------------------------------------------------------------------------
\subsection{Hydrodynamics and effective binary mixtures}
\label{subsec:average}
%---------------------------------------------------------------------------------------------------
The PIJ model can handle more than three components in a mixture. In these cases, it provides $M(M-1)/2$ independent mutual diffusion coefficients $D_{ij}$ with $i,j = 1 \dots M$ and $i\neq j$, that appear in the multi-species Navier-Stokes (NS) equations. Of particular interest are the continuity equations \cite{SIMA16} that we write in the absence of temperature and pressure gradients as
\begin{subequations}
\begin{equation}
\partial_t (\rho_i) + \nabla \cdot \left( \rho_i \textbf{u} + \textbf{F}_i \right) = 0,
\end{equation}
\begin{equation}
\label{closure_diffusion}
\textbf{F}_i = -\rho_i \, \sum_{j=1}^M D_{ij} \, \nabla x_j,
\end{equation}
\begin{equation}
\label{Dij_sym}
D_{ji} = D_{ij},
\end{equation}
\begin{equation}
\label{Dij_sum}
\rho_i\,D_{ii} = -\sum_{j=1;\,j\neq i}^M \rho_j\,D_{ji},
\end{equation}
\end{subequations}
with $\rho_i$ the mass density of species $i$, $\textbf{u}$ the mass average velocity, $\textbf{F}_i$ the diffusive mass flux of species $i$, and $x_i$ the number density fraction of species $i$.

Equations\,\eqref{Dij_sym} and \eqref{Dij_sum} imply that, in the case of binary mixtures, from four coefficients, one gets only one, $D_{12}$, while in the case of ternary mixtures, from nine coefficients, the problem reduces to three, $D_{12}$, $D_{13}$, and $D_{23}$ and so on. Comparing the interdiffusion process between binary and ternary mixtures, the phenomenology is far more rich and complex in the latter with possible upstream diffusion for instance. Adding more species than in the ternary mixtures does not introduce any physically new phenomena.

We generally consider microscopic ternary mixtures to contain three distinct species.  However, as is often the case in applications,  two components can form a unique material, e. g., CH, which is the ablator of ICF capsules.

%************************************** figure***************************************************************************************
\begin{figure}[t]
\begin{center}
\includegraphics[scale=0.4]{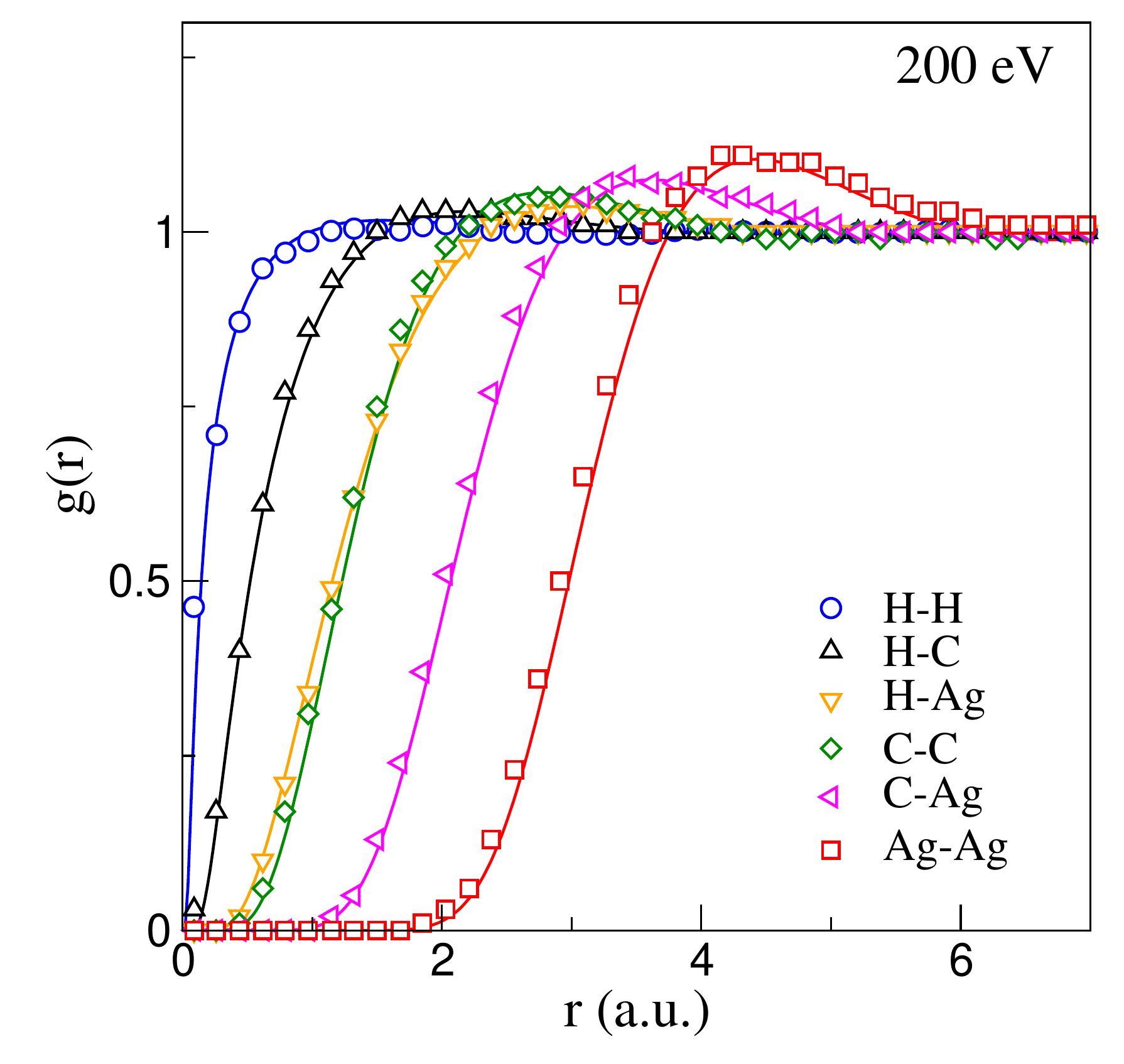}
\caption{(Color online) Comparison of all pair correlation functions obtained by OFMD simulations (symbols) with MCHNC calculations (solid lines) in a HCAg mixture at 200\,eV and 720\,Mbar. Self-correlations: $g_\text{HH}$: blue circles; $g_\text{CC}$: green diamonds; $g_\text{AgAg}$: red squares. Cross correlations: $g_\text{HC}$:  black up triangles; $g_\text{HAg}$: orange down triangles; $g_\text{CAg}$: indigo left triangles.
\label{fig:allcorr} }
\end{center}
\end{figure}
%**************************************figure***************************************************************************************

%**************************************big figure***************************************************************************************
\begin{figure*}[]
\begin{center}
\includegraphics[scale=0.65]{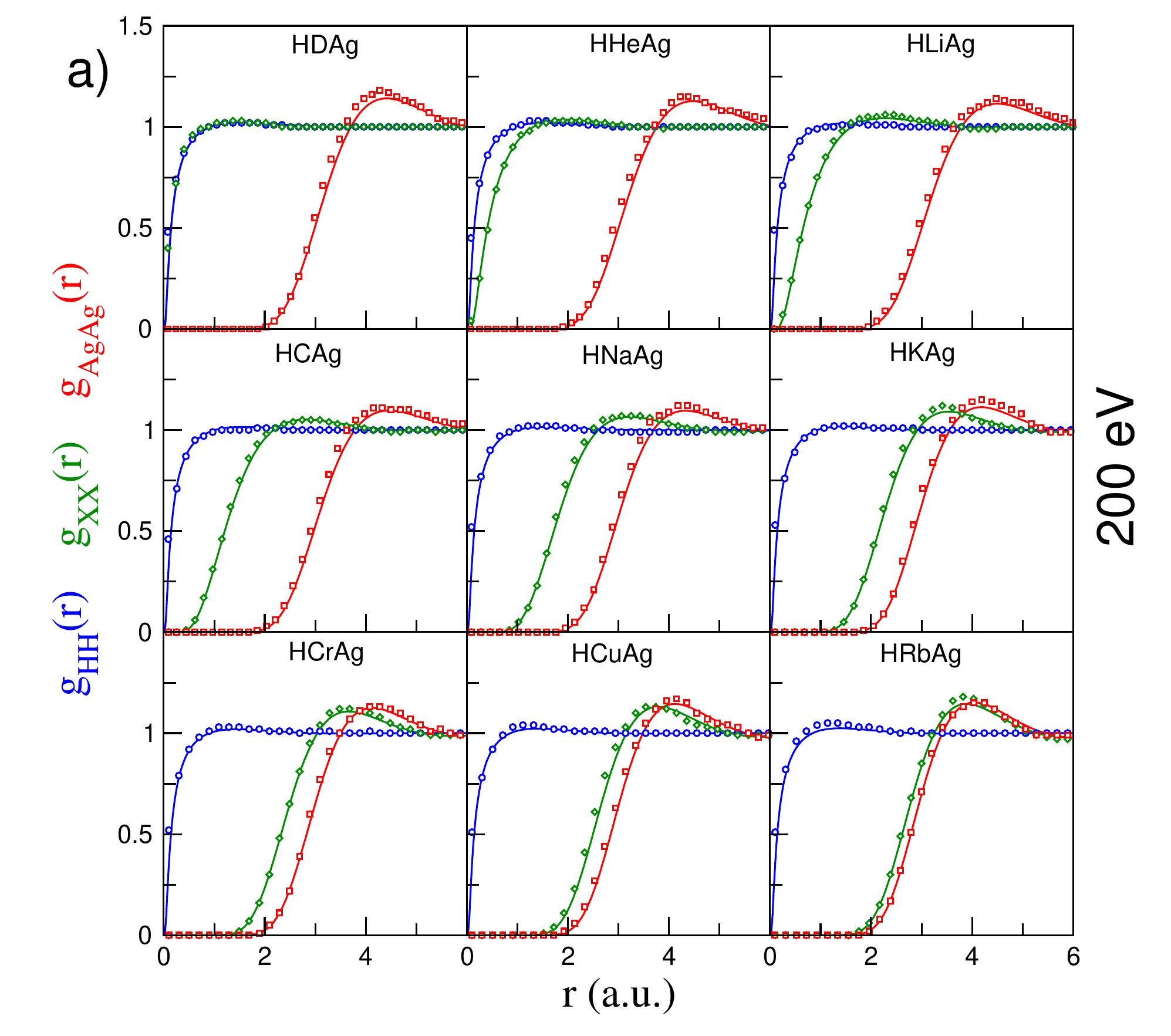}
\includegraphics[scale=0.65]{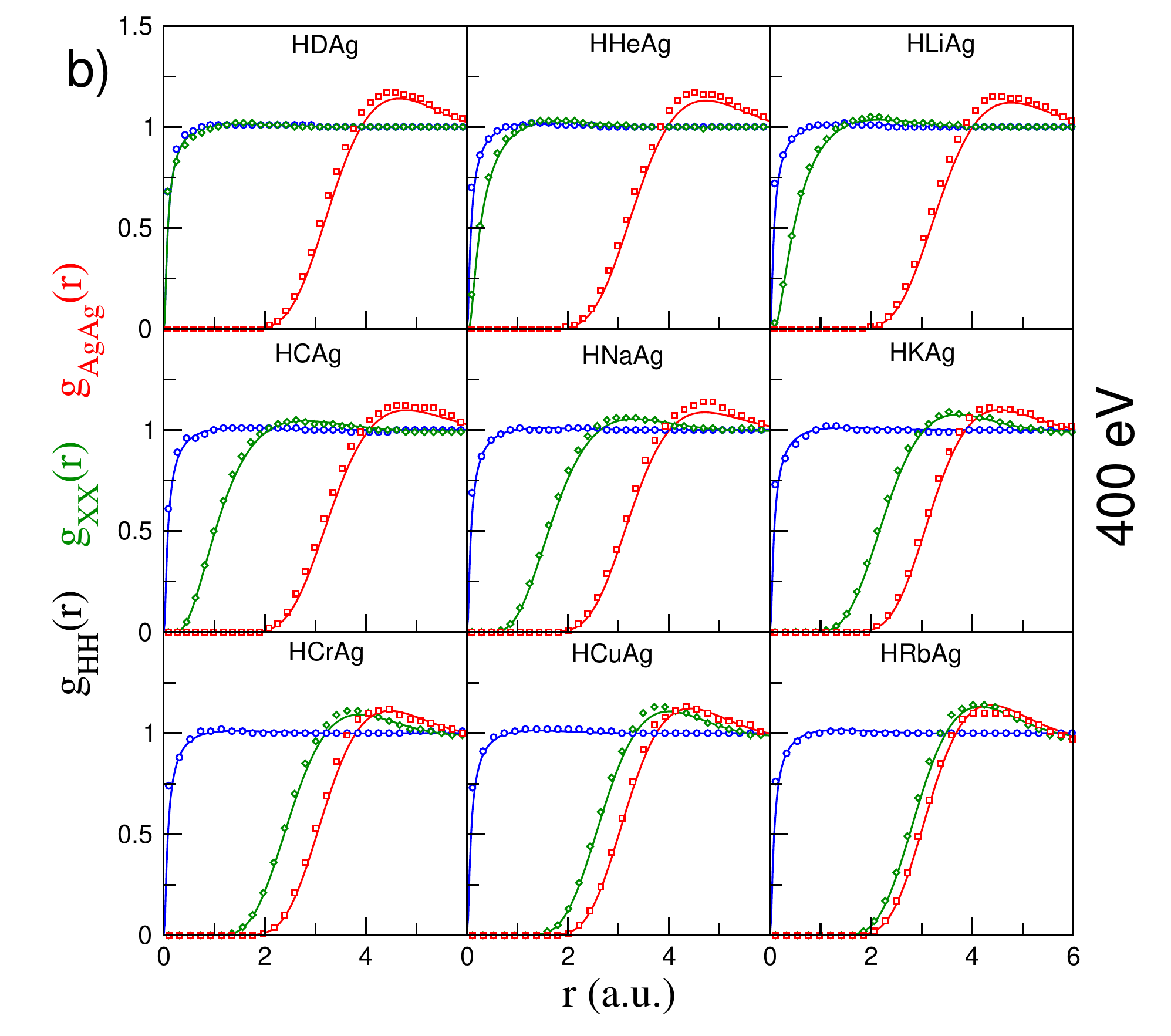}
\caption{(Color online) Pair distribution functions  for various H-X-Ag mixtures (X=D, He, Li, C, Na, K, Cr, Cu and Rb)  at 200~eV and 720\,Mbar(a) and 400~eV and 1520\,Mbar (b) and for proportions (0.4: 0.4: 0.2) in number. Symbols for OFMD simulations are: blue circles for $g_\text{HH}$; green diamonds for $g_\text{XX}$ and red squares for $g_\text{AgAg}$. Corresponding solid lines are the MCHNC calculations using the\iso\, rule. 
\label{fig:gall200} } 
\end{center}
\end{figure*}
%**************************************big figure***************************************************************************************

We here derive a general formula to define the mutual diffusion of an effective binary mixture starting from the set of corresponding coefficients of a multicomponent system. Assume that the components are labeled in such an order that the indices $i$ from 1 to $L$ correspond to one effective species denoted $E_1$ and the other indices from $L+1$ to $M$ correspond to the other effective species denoted $E_2$. They evolved as a whole and keep the same partition defined by
\begin{equation*}
\alpha_i= 
\left\lbrace
\begin{aligned}
&\frac{x_i}{X}, \hspace{0.5cm} \text{if} \hspace{0.2cm} i\in E_1,\\
&\frac{x_i}{1-X}, \hspace{0.2cm} \text{if} \hspace{0.2cm} i\in E_2,
\end{aligned}
\right.
\end{equation*}
with $X=\sum\limits_{j=1}^{L}x_j$. Since the $\alpha_i$ are constant coefficients, they share the same gradient 
\begin{equation*}
\nabla x_i= 
\left\lbrace
\begin{aligned}
&\alpha_i \nabla X, \hspace{0.5cm} \text{if} \hspace{0.2cm} i\in E_1,\\
&-\alpha_i \nabla X, \hspace{0.2cm} \text{if} \hspace{0.2cm} i\in E_2,
\end{aligned}
\right.
\end{equation*}
Adding the first continuity equations, from $i=1$ to $L$, corresponding to the set $E_1$, leads to
\begin{equation}
\label{eq_E1}
\partial_t ( \rho^{E_1} ) + \nabla.( \rho^{E_1} \textbf{u} + \textbf{F}^{E_1})=0, 
\end{equation}
where $\rho^{E_1}$ and $\textbf{F}^{E_1}$ are defined by 
$\rho^{E_1} = \sum_{i=1}^L \rho_i$, 
and $\textbf{F}^{E_1} = \sum_{i=1}^L \textbf{F}_i$. Similar relations are found for the quantities $\rho^{E_2}$ and $\textbf{F}^{E_2}$ summing over $i$ from $L+1$ to $M$. Now combining (\ref{closure_diffusion}) with the previous relations leads to the following closure
\begin{align*}
\textbf{F}_i &= - \Big(\sum_{j \in E_1} \rho_i D_{ij} \alpha_j - \sum_{j \in E_2} \rho_i D_{ij} \alpha_j \Big) \nabla X,\\
&= - \rho \Big( \sum_{j \in E_1} y_i D_{ij} \alpha_j - \sum_{j \in E_2} y_i D_{ij} \alpha_j \Big) \nabla X.
\end{align*}
where $y_i = \rho_i / \rho$ are the mass fractions. One recovers the form used for binary mixture with an effective diffusion coefficient \Dh
\begin{equation*}
\textbf{F}^{E_1}=-\rho \Dh \nabla X,
\end{equation*}
where
\begin{equation}
\Dh=\sum\limits_{i \in E_1}\left( \sum\limits_{j \in E_1} y_i D_{ij} \alpha_j - \sum\limits_{j \in E_2} y_i D_{ij} \alpha_j \right). 
\label{eq:effective}
\end{equation}

Eq.~(\ref{eq:effective}) gives the hydrodynamical definition of the mutual diffusion between two effective species $E_1$ and ${E_2}$. It is to be compared with an empirical binary mixture of average elements. For example, an  equimolar CH mixture is often treated as an effective material of  atomic number Z=3.5 and atomic mass A=6.5. Is this reduction to an effective binary mixture ([CH]-Ag) representative of the full mixture? Is the integrity of the initial compound of CH conserved when microscopic diffusion affects differently its components (due to asymmetry of mass and charge, leading to extremely different coupling and behavior with respect to diffusion)\,? The PIJ model, which can handle mixtures with all components or with effective components, can be used to address these questions. 

Eq.~(\ref{eq:effective}) is given explicitly in Appendix \ref{app:hydro} for three  and four components by (\ref{eq:3eff})  and (\ref{eq:4eff}), respectively.
%---------------------------------------------------------------------------------------------------
\section{Structure}
\label{sec:structure}
%---------------------------------------------------------------------------------------------------
Using the MCHNC approach  with Coulomb interactions, we test the ability of the  \iso\, prescription at giving adequate ionizations to predict the static structure for various mixtures H-X-Ag with X=D, He, Li, C, Na, K, Cr, Cu and Rb  at 200 eV and 400~eV. The proportions for each species are (0.4:0.4:0.2) in number. 
Tables~\ref{table2} and \ref{table3} give  the total density, the Wigner-Seitz radius $a$, the ionic radii, the charges, and the coupling parameters under the  \iso\, prescription.

To mimic a mixing layer, the different mixtures are held at the same pressure for each temperature by adjusting the density of the mixture. Since the pressure in this temperature regime is dominated by the electronic contribution, this constraint is very close to the \iso\, constraint. Therefore, the ionizations and the coupling parameters of H and Ag do not change significantly when substituting the middle element (see Table~\ref{table2}). Hydrogen is clearly in the kinetic regime since its coupling parameter is smaller than unity and decreases by almost a factor of two when the temperature doubles from 200 to 400~eV which means that H is fully ionized. In contrast, the coupling of Ag, which is larger than 10,  does not change much with the temperature, due to the $\Gamma$-plateau, where ionization (squared) is balanced by the temperature.

%---------------------------------------------------------------------------------------------------
\subsection{Correlations}
%---------------------------------------------------------------------------------------------------
We recall that for $M$ species there are $M(M+1)/2$ PDFs; $M$ self correlations and $M(M-1)/2$ cross correlations.  For a HXAg mixture we have three self PDF ($g_\text{H-H}$, $g_\text{X-X}$,  $g_\text{Ag-Ag}$ ) and three cross PDF ($g_\text{H-X}$, $g_\text{X-Ag}$ and $g_\text{H-Ag}$ ). We show in Fig.\,\ref{fig:allcorr} all correlations obtained with an OFMD simulation of a ternary mixture of HCAg at 200\,eV. The symbols used throughout the paper  for self correlations are: blue circles for $g_\text{HH}$;  green diamonds for $g_\text{CC}$ and  red squares for $g_\text{AgAg}$. For  cross correlations, we use   black up triangles for $g_\text{HC}$; orange down triangles for $g_\text{HAg}$ and  indigo left triangles for $g_\text{CAg}$. The agreement with the MCHNC calculations (solid lines) is excellent for all correlations.  

In the following we will ignore the cross-correlations in the figures for the sake of clarity.

%---------------------------------------------------------------------------------------------------
\subsection{Global comparison}
%---------------------------------------------------------------------------------------------------
%Fig.~\ref{fig:gall200} shows the corresponding PDFs.  Symbols for OFMD simulations are: blue circles for $g_\text{HH}$;  green diamonds for $g_\text{XX}$ and red squares: $g_\text{AgAg}$.  MCHNC calculations using data given in Tables~\ref{table2} and \ref{table3} are shown by corresponding solid lines. 
Mixtures shown in Fig.~\ref{fig:gall200}(a) for 200\,eV span all possible combinations of the MZE. HDAg shows no distinction between H and D, as expected. HCAg displays an evenly structured mixture, and HRbAg is a very asymmetric mixture with PDFs that peak near the same distance. The MCHNC  predicts PDFs in close agreement with the simulations for all configurations. A similar agreement is obtained for 400~eV  as shown in Fig.~\ref{fig:gall200}b. 
Interestingly for all mixtures,   the position of the Ag peak appears relatively independent of temperature. This is a manifestation of the "$\Gamma$-plateau" for which the reduced density, $\rho_{c}=\rho/(ZA)$,  equals 0.0033, being less than 0.0045 \cite{ARNA13}.

As already observed for binary mixtures, the structure at both temperatures confirms the coexistence of a purely kinetic species with an almost flat PDF %\cite{comment2}%
 (hydrogen in blue), with a strongly coupled species, characterized by a well defined peak (Ag in red). 

The structure of mixtures with neighboring light components such as HDAg, HHeAg, HLiAg or neighboring  heavy components such as HCuAg or HRbAg raises the question of replacing two components by a single effective component. This question will be addressed later in Sec.~\ref{subsec:average}.

%---------------------------------------------------------------------------------------------------
\subsection{Detailed analysis of HCAg structure}
\begin{figure}[]
\begin{center}
\includegraphics[scale=0.4]{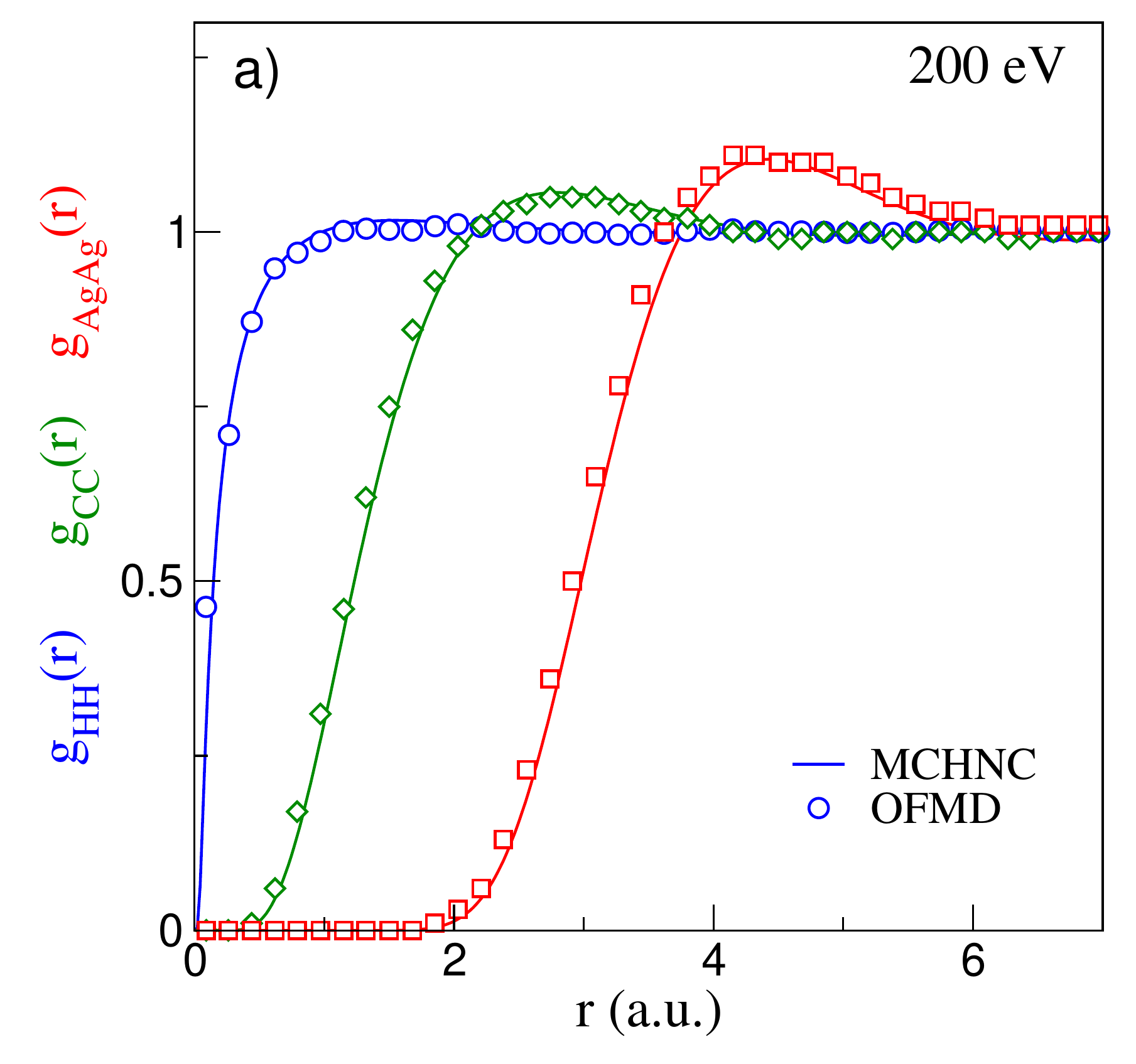}
\includegraphics[scale=0.4]{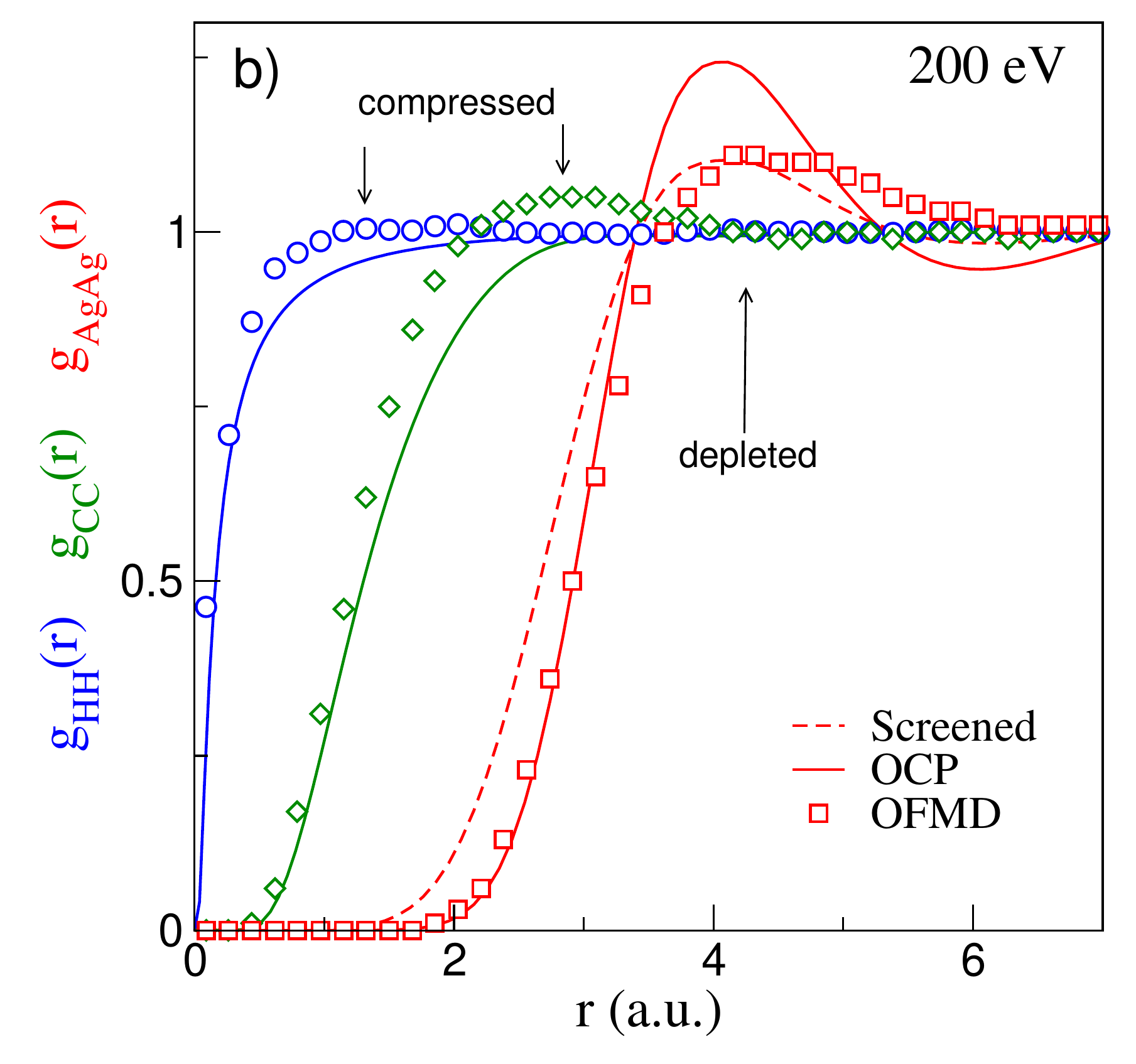}
\caption{(Color online) Same symbols as in Fig.\,\ref{fig:gall200}. a): Pair distribution functions for a  HCAg mixture. Symbols are OFMD simulations and solid lines MCHNC calculation using the \iso\ rule. b): comparison of OFMD results (symbols) with the OCP structure (solid lines) using coupling parameters given in Table~\ref{table2} ($\Gamma_\text{HH}=0.13$, $\Gamma_\text{CC}=1.88$  and $\Gamma_\text{AgAg}=14.7$) and with the screened formula (\ref{lab:correc1}) (red dashed line).
 \label{fig:fits} } 
\end{center}
\end{figure}

%---------------------------------------------------------------------------------------------------
We detail in Fig.~\ref{fig:fits}  the structure of the HCAg mixture at 200~eV to emphasize the efficacy of the \iso\, approach coupled with the MCHNC calculation.  Circles are for the OFMD simulation of the whole mixture with symbols in blue for hydrogen, green for carbon, and red for silver. We recall in Fig.~\ref{fig:fits}a, that  MCHNC  with the charges $\{Q_i\}$ given by the \iso\, prescription  accurately reproduces the structure of the mixture.    Now, instead of using  the charges,  we use the coupling parameters $\{\Gamma_i\}$, given in Table~\ref{table2}, to try to establish for each component a connection with the  OCP structure at the corresponding coupling.   We observe  in  Fig.~\ref{fig:fits}b that the OCP PDF (solid red line) predicts a much higher peak of Ag than OFMD simulations, and the OCP PDFs for hydrogen (solid blue line) and carbon (solid green line) fall well below the corresponding OFMD data. In other words, hydrogen and carbon are compressed and silver depleted. If we use the screening correction given by Eq.~(\ref{lab:correc1}) with $x_\text{HZE}=0.2$, giving $\alpha=0.8667$ and $\Gamma_\text{HZE}=8.262$ (instead of $14.7$),  we restore the intensity of the $g_\text{AgAg}$ peak, but we do not get the correct excluded volume. The same conclusions for the impossibility of obtaining the structure from one-component theories could be drawn for hydrogen and carbon. A multicomponent approach  is definitively  necessary to accurately reproduce the structure of the mixture. 
%**************************************   figure***************************************************************************************
%******************************************************************************************************************************************

%---------------------------------------------------------------------------------------------------
\section{Transport coefficients}
%---------------------------------------------------------------------------------------------------
The \iso\, prescription, tested with respect to static structure in Sec.\,\ref{sec:structure}, is now used in the PIJ model for transport coefficients.

 OFMD simulations of the HCAg mixture with a varying concentration in HZE (Ag) ((1-x)/2:(1-x)/2:x) were performed at temperatures of 100, 200 and 400~eV \cite{WHIT19}. 
 %Self diffusion and  viscosity coefficients were computed using analytical fits of the measured autocorrelation functions as described in \cite{WHIT19}. 
 The mutual diffusion coefficients were deduced from self-diffusion coefficients using Darken approximations.
 The charges and coupling parameters predicted by the \iso\, prescription are reported in Table~\ref{table4} for 200~eV. For a given temperature,  the ionizations and the ionic radii are barely affected by the concentration in the heavy element since the total density is adjusted to keep the pressure constant, as in Sec.~\ref{sec:structure}.\\

%**************************************      Triple figure***************************************************************************************
\begin{figure*}[]
\begin{center}
\includegraphics[scale=0.65]{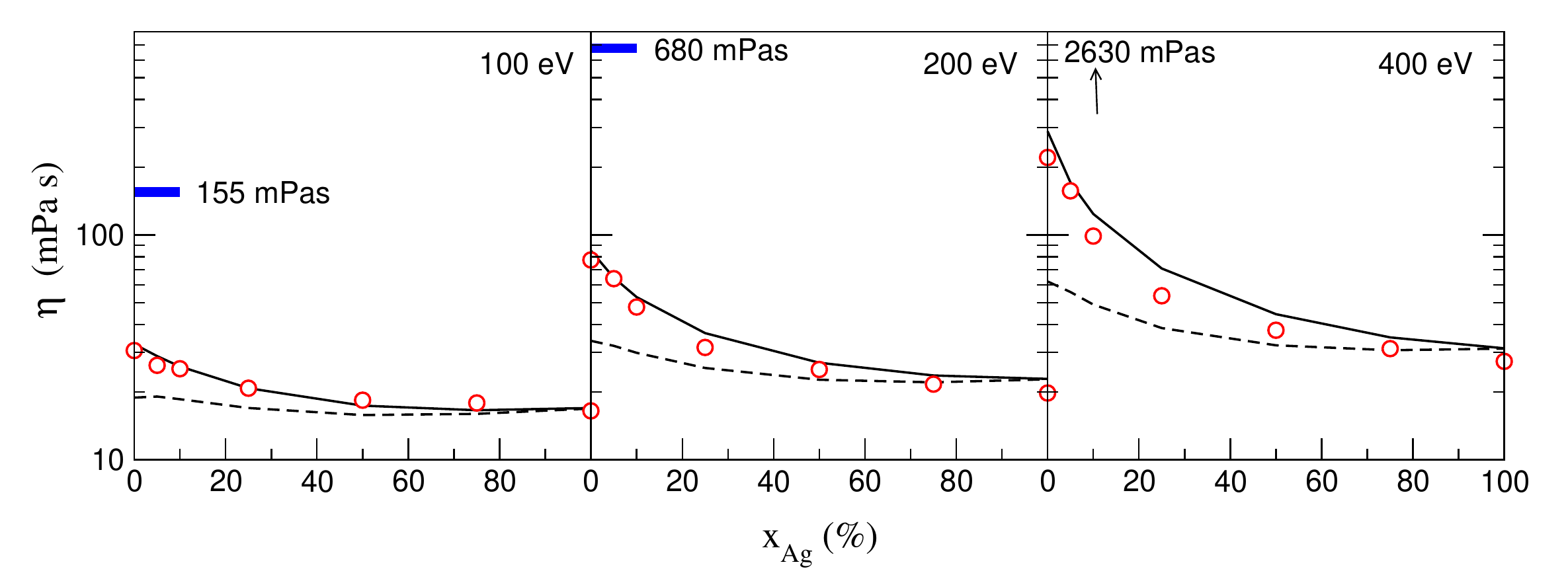}
\caption {\label{visco}  Color online)Viscosity coefficient $\eta$  of an equimolar hydrogen-carbon (HC) mixture   in the presence of an increasing proportion of silver (Ag). Circles are  OFMD simulations results \cite{WHIT19} and solid lines are the PIJ estimations. The dashed lines represent the OCP-related excess contributions in the PIJ calculations. The viscosity of pure hydrogen is indicated by the blue bars (off-scale for 400\,eV). }
\vspace{1 cm}
\includegraphics[scale=0.65]{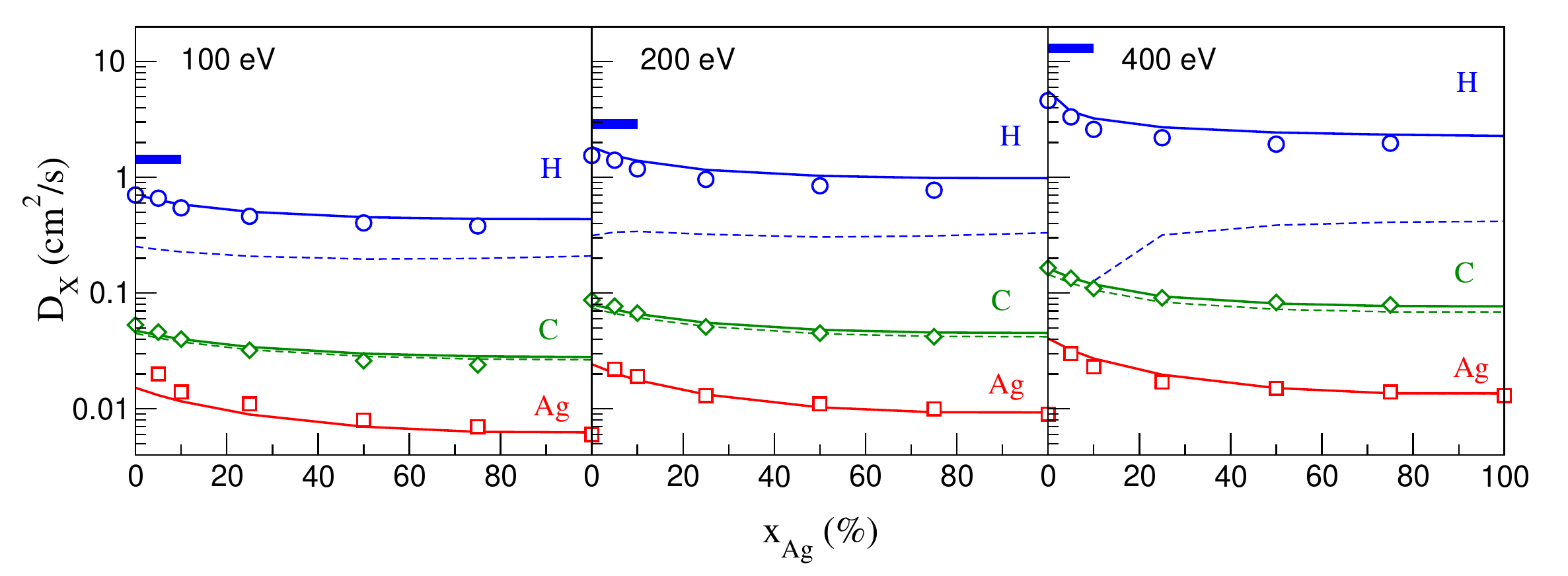}
\caption {\label{self} (Color online) Self-diffusion coefficients of H (blue circles), C (green diamonds), and Ag (red squares) as a function of  silver (Ag) concentration, added to an equimolar hydrogen-carbon (HC) mixture. Symbols are for OFMD simulations and lines for PIJ model.
The dashed lines represent the non-negligible OCP-related excess contributions in the PIJ calculations, dominant for Ag. The self diffusion of pure hydrogen is indicated by the blue bars. }
\vspace{1 cm}.\includegraphics[scale=0.65]{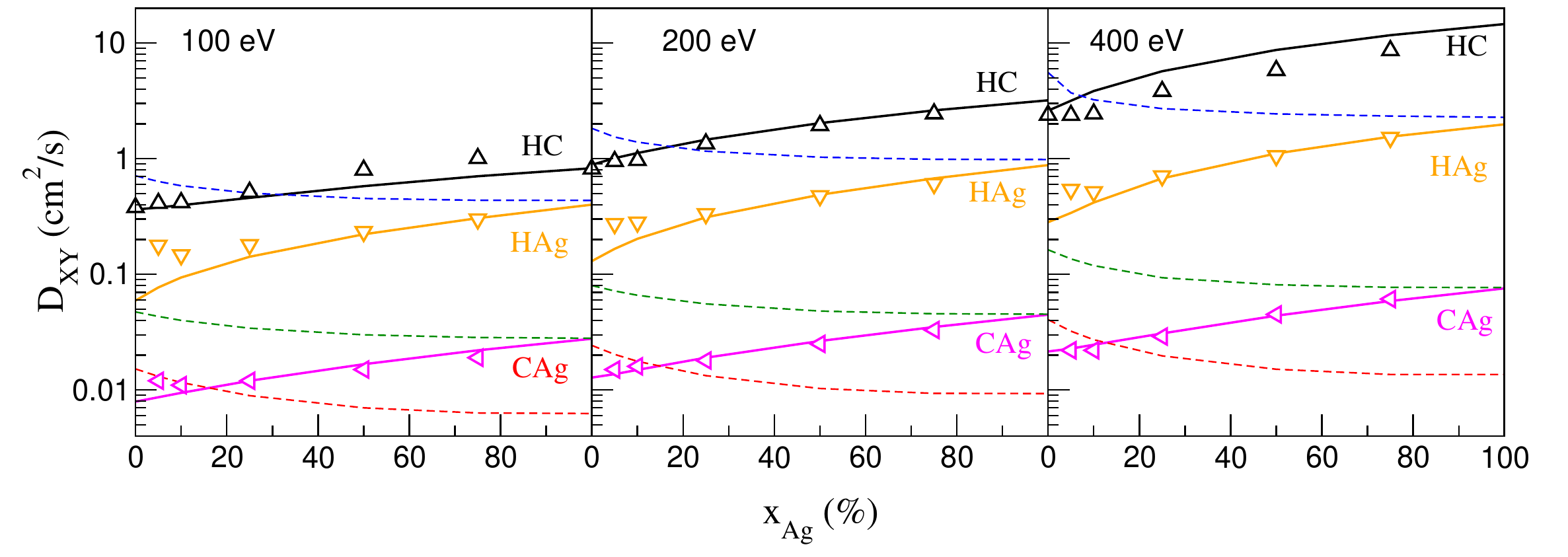}
\caption {\label{mutual}(Color online) Mutual diffusion coefficients $D_\text{HC}$ (black up triangles), $D_\text{HAg}$ (orange down triangles) and $D_\text{CAg}$ (magenta left triangles).  Dashed lines are previous self diffusion constants of hydrogen (up dashed blue line), carbon (middle dashed green line) and silver (bottom dashed red line), shown in Fig.\,\ref{self}.}
\end{center}
\end{figure*}
%**************************************big figure***************************************************************************************

%---------------------------------------------------------------------------------------------------
\subsection{Viscosity}
%---------------------------------------------------------------------------------------------------
 In Fig.\,\ref{visco}, the OFMD results for viscosity are compared with PIJ estimations as a function of Ag atomic fraction $x_\text{Ag}$ for 100, 200, and 400\,eV. The agreement is fairly good with deviations less than 10\% at 100\,eV, 20\% at 200\,eV, and 25\% at 400\,eV. 

The excess contributions correcting the kinetic calculation of PIJ at strong coupling with OCP-related estimations are shown as dashed lines. Theses corrections represent the potential contributions to the viscosity arising from the autocorrelation of the stress tensor, Eq.\,\eqref{STACF}.  In the binary system of CH, when $x_\text{Ag} = 0$, these corrections amount to 60\% of the viscosity at 100\,eV, 50\% at 200\,eV, but only 20\% at 400\,eV. As the Ag concentration increases, these corrections contribute more to the viscosity and become dominant for pure silver. 

This balance between kinetic and potential contributions explains why the viscosity increases with temperature more strongly in the binary CH mixture than in the pure Ag system. The combined effect of temperature and concentration variations gives rise to a reduction of viscosity by an order of magnitude when $x_\text{Ag}$ varies from 0 to 1 at 400\,eV.

For the sake of comparison, we have also computed with PIJ the viscosity of pure hydrogen (without C and Ag) at the same pressure and temperature. We obtain $155$, $680$, and $2630$~mPa\,s for 100, 200, and 400~eV (blue bars  in Fig.\,\ref{visco}), well beyond the corresponding values of hydrogen mixed with carbon. This explains the relatively smooth reduction of the viscosity with an increasing addition of silver in CH. The strongest  effect is already caused by the CH mixing itself.

%---------------------------------------------------------------------------------------------------
\subsection{Self diffusion}
%---------------------------------------------------------------------------------------------------
In Fig.\,\ref{self}, the OFMD results for self-diffusion coefficients are compared with PIJ estimations as a function of Ag atomic fraction $x_\text{Ag}$ for 100, 200, and 400\,eV. The agreement is fairly good with deviations less than 20-30\%. The dashed lines represent the excess OCP-related contributions to the PIJ calculations as for Fig.\,\ref{visco}. These contributions are dominant for the C and Ag self diffusions in all cases, and negligible for H self diffusion except at 100\,eV where they account for 40-60\%. 

To better understand the phenomenon at work, we detail the different situations corresponding to the same pressure and  temperature.

We first consider pure hydrogen. In the lack of carbon and silver, a PIJ estimation of the self diffusion of hydrogen gives $1.4, 2.9$, and $13\,\text{cm}^2/\text{s}$ at  respectively 100, 200 and 400~eV (tick blue segment on the $y$-axis).

Second, in a binary mixture of hydrogen and carbon, the hydrogen self diffusion is reduced by the presence of 50\% of carbon to $0.71$, $1.8$, and $5.6 \,\text{cm}^2/\text{s}$ for corresponding temperatures.

Third, in the ternary mixture of equimolar carbon-hydrogen mixture with increasing proportion of a heavy material, the self diffusion of hydrogen and carbon are continuously reduced by the amount of silver. %No dramatic drop is observed as for binary mixtures. Self diffusion coefficients are well reproduced by PIJ model (corresponding dashed lines).

%---------------------------------------------------------------------------------------------------
\subsection{Mutual diffusion}
%---------------------------------------------------------------------------------------------------
In Fig.\,\ref{mutual}, the OFMD results for mutual diffusion are compared with PIJ  estimations as a function of Ag concentration $x_\text{Ag}$ for 100, 200 and 400~eV.    The agreement is satisfying with deviations less than around 20-30\%, except at low concentration of Ag where $D_\text{HAg}$ is overestimated by PIJ by a factor of 2.

The simplest case to interpret is the CAg mutual diffusion, because in this case, H can be neglected and the mixture reduced to a binary CAg mixture, with a varying proportion of Ag. We have studied similar cases \cite{TICK16}, and we recognize the behavior of the mutual diffusion, approximately interpolating between the Ag and C self-diffusion coefficients (shown by corresponding dashed lines). Limits are not exact since we have neglected H, but PIJ recovers this "Z shaped" behavior very precisely. Since the self-diffusions $D_\text{C}$ and $D_\text{Ag}$ are dominated by the OCP contribution (dashed lines in Fig.\,\ref{self}), it is not astonishing to recover the Darken relations.

The mutual diffusion between H and C (solid blue line) increases also with the Ag concentration (blue circles), but is almost always bigger than corresponding hydrogen self-diffusion in the mixture (dashed blue line). This is at variance with the usual Darken relation in binary mixtures. The reformulation of the Darken approximation in multicomponent mixtures given in the Appendix by  (\ref{app:darken}) clearly shows that if  $D_\text{Ag}$ is much smaller than  $D_\text{C}$, we always obtain a HC mutual diffusion higher than H and C self diffusion (in the mixture). 
Note that the mutual diffusion between hydrogen and carbon at any concentration always stays below pure hydrogen self diffusion  in the same conditions of temperature and pressure (bar on $y$-axis).

The behavior of the HAg mutual diffusion coefficient with Ag concentration, which follows the multicomponent Darken relation, is also reproduced by the kinetic contribution that dominates the PIJ estimation.

%***************************************** double Figure ***************************************************************************************************
\begin{figure}[t]
\begin{center}
\includegraphics[scale=0.6]{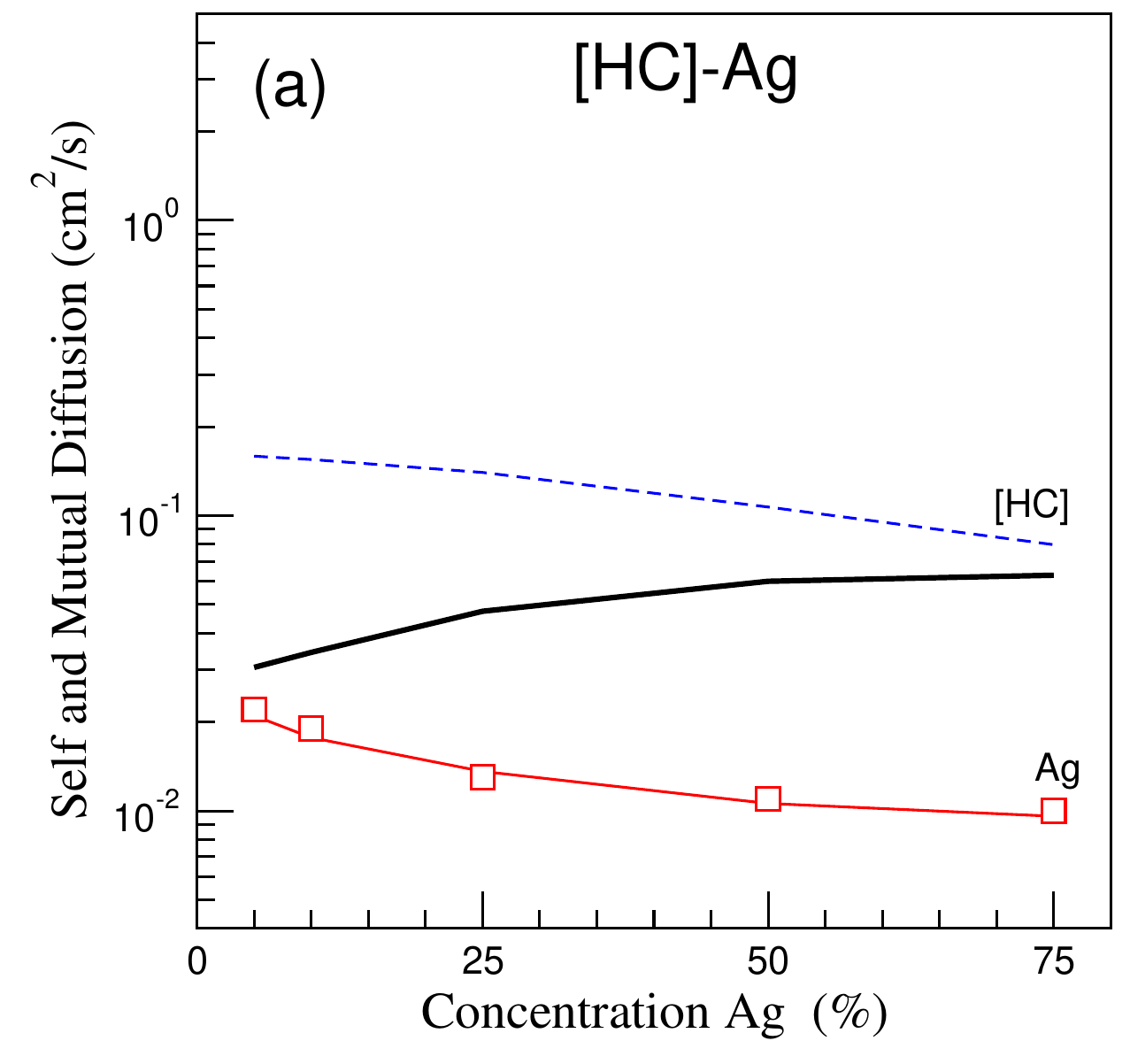}
\includegraphics[scale=0.6]{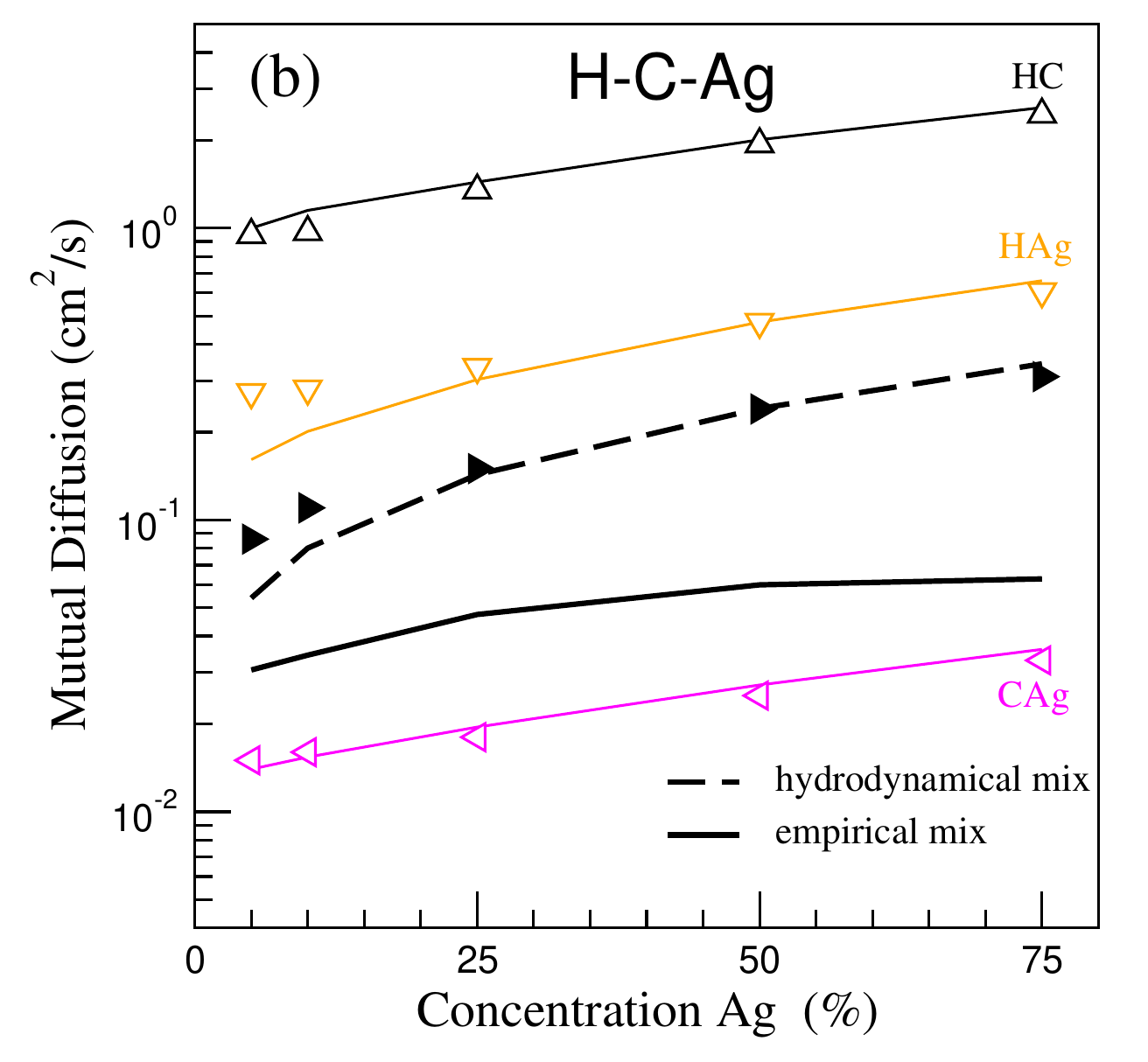}
\caption {\label{fig:effective}(Color online) Evaluations of self and mutual diffusion coefficients of a H-C-Ag mixture  at 200~eV. (a) Self and mutual diffusion coefficients in the empirical mix. Blue dashed line: self diffusion of [CH], red solid line:  self diffusion of Ag, and heavy solid black line: mutual diffusion \De\:of the empirical binary mixture [CH]-Ag. (b) Mutual diffusion coefficients in the ternary H-C-Ag mixture. OFMD results (black up triangles: HC; orange down triangles: HAg, and magenta left triangles CAg). Lines of corresponding colors  are the PIJ evaluations.  In dashed black, the hydrodynamical mix \Dh\,when computed with PIJ (dashed line), or  with OFMD data (solid black triangles). The solid black line is the previous \De\,of Fig.\:\ref{fig:effective}(a) for comparison.}
\end{center}
\end{figure}
%*****************************************************************************************************************************************************************

%---------------------------------------------------------------------------------------------------
\subsection{Effective binary mixture}
\label{sec:effective}
%---------------------------------------------------------------------------------------------------
 We address next the question of finding an effective binary mixture for a multicomponent mixture ($M\ge 3$). For a H-C-Ag mixture it seems natural to consider a single  [CH] component with atomic number 3.5 and an effective mass of 6.5 for a 50\% composition. The mutual diffusion of the empirical mixture (solid black line), obtained by using PIJ with two components: [CH] and Ag, is plotted in Fig.~\ref{fig:effective} (a). It interpolates between the self diffusion of Ag (red solid line) at vanishing concentration and the [HC] self diffusion (blue dashed line) for vanishing [HC]  concentration, in agreement with the Darken relation for the effective [HC]-Ag mixture.
 
 We consider the whole H-C-Ag mixture and compute all mutual diffusion coefficients as shown in Fig.~\ref{fig:effective} (b). The OFMD results are represented by black up triangles for HC; orange down triangles for HAg, and magenta left triangles  for CAg. Solid lines of corresponding colors  are the PIJ evaluations. An effective diffusion coefficient in the hydrodynamic limit is defined  in subsection \ref{subsec:average}, whose explicit formulation for three components is given  by (\ref{eq:3eff}) in the Appendix (hydrodynamical mix).  The hydrodynamical mix can be obtained either from the OFMD simulation results (solid black triangles), or by  PIJ (black dashed line). Both calculations are in excellent agreement. This evaluation is a few times higher than the empirical binary mixture and raises questions as to  the validity of the empirical mix commonly used. The hydrodynamical mix yields characteristic times of mixing much lower than empirical mixing evaluations.
\begin{table}[b]
\caption{Mutual diffusion and mixing times  for empirical binary mixture \De\:and effective hydrodynamical binary mixtures \Dh\:for 3 and 4 components mixtures. The averaged effective component is shown by  brackets : [CH], [DT]. Note the change of time unit for the [CH][DT] and  [CH]He mixtures which is in ps. $x$ is the proportion of Au in the Au mixture, of [DT] in the [CH]-[DT] mixture and He in the [CH]-He mixture.}
\begin{center}
\begin{tabular}{|c|c|c|c|c|c|c|c|c|}
\hline
\multicolumn{9}{|c|}{[CH]Au}	\\
\hline
$x$	&	$T$	&	$\rho$	&	$Q_\text{[CH]}$	&	$Q_\text{Au}$	&	\De		& $\tau$ & \Dh & $\tau^{\text{eff}}$ 	\\
\%	&	eV	&	\udens	&			&			&	cm$^2$/s		&	ns	& cm$^2$/s & ns\\
\hline
10		& 100	&	5			&	2.9		&	14		&	{\bf 3.0 10$^{-2}$}	&	330 			& {\bf 8.7 10$^{-2}$}	 	& 115	\\
10		& 200	&	5			&	3.2		&	21		&	{\bf 5.4 10$^{-2}$}	&	185			& {\bf 15. 10$^{-2}$} 	& 66		\\
10		& 500	&	5			&	3.4		&	37		&	{\bf 0.12}			&	83			& 	{\bf 0.45} 			& 22		\\
10		& 1000	&	5			&	3.5		&	50		&	{\bf 0.21}			&	48			& 	{\bf 1.3} 			& 7.7		\\
10		& 1000	&	10$^{-2}$		&	3.5		&	64		&	{\bf 32}			&	0.31 			& 	{\bf 379} 			& 0.03 	\\
\hline
\hline
\multicolumn{9}{|c|}{[CH][DT]}	\\
\hline
$x$	&	$T$	&	$\rho$			&	$Q_\text{[CH]}$	&	$Q_\text{[DT]}$	&	\De		& $\tau$ & 	\Dh	& $\tau^{\text{eff}}$ \\
\%	&	eV	&	\udens			&			&			&	cm$^2$/s		&	ps	&	cm$^2$/s & ps  \\
\hline
10		& 1000	&	20			&	3.4		&	1		&	{\bf 32}	&	310 & {\bf 37} & 270			\\
50		& 1000	&	20			&	3.4		&	1		&	{\bf 22}	&	450 & {\bf 20} & 500			\\
\hline
\hline
\multicolumn{9}{|c|}{[CH]He}	\\
\hline
$x$	&	$T$	&	$\rho$			&	$Q_\text{[CH]}$	&	$Q_\text{He}$	&	\De	& $\tau$ & \Dh	& $\tau^{\text{eff}}$\\
\%	&	eV	&	\udens			&			&			&	cm$^2$/s		&	ps	&	cm$^2$/s & ps\\
\hline
50		& 200	&	5 10$^{-3}$	&	3.5		&	2		&	{\bf 39}		&	260 & {\bf 163}		& 60	\\
50		& 1000	&	5 10$^{-3}$	&	3.5		&	2		&	{\bf 1.1 10$^3$}	&	9 & {\bf 5.4 10$^3$} 	& 1.9\\
\hline

\end{tabular}
\end{center}
\label{bin}
\end{table}%

%---------------------------------------------------------------------------------------------------
\section{ICF mixtures}
%---------------------------------------------------------------------------------------------------
The PIJ model is now used to predict characteristic mixing times or lengths related to ICF situations. One must distinguish evaluations of mixing times of two neighboring materials (CH with DT, Au with CH) leading to an homogeneous microscopic mixing, and evaluations of particle escape in an ICF target (Fig.1) from a material in the case of "separated reactants" experiments. In the first case we need the mutual diffusion evaluations and in the second case the self diffusion coefficient of the escaping particle.
%---------------------------------------------------------------------------------------------------
\subsection{ Mix in hohlraums and capsules}
%---------------------------------------------------------------------------------------------------

We first consider the mixing of CH ablator  with helium atoms of the gas filling the hohlraum. Densities are of the order of a few 10$^{-3}$\udens\, for CH and  10$^{-4}$\udens\, for He. Temperatures range from a few hundreds eV to a few keV. In these conditions, helium is fully ionized and weakly coupled, while CH is ionized 3 times and moderately coupled. The He-Au mixing is more asymmetric with gold ionized 40-50 times and thus strongly coupled ($\Gamma \simeq 5$).
Another interesting mixture is CH  ablator mixed with DT inside the capsule.

We shall give order-of-magnitude estimates for these mixing situations using scaling arguments.
With a mutual diffusion coefficient $D$, a material of L=1\,$\mu$m of thickness will be mixed in a characteristic time  $\tau \simeq L^2/D = 10$/D(\text{cm$^2$/s}) ns (1\,cm$^2$/s = 0.1\,$\mu \text{m}^2/ \text{ns}$). A mutual diffusion of $10^{-1}$ cm$^2$/s means that mixing will occur in 100~ns, which is of the same order as observed in hydrodynamics calculations. Table \ref{bin} gives orders of magnitude of mixing time for a one micron layer of CH. One can see that the mixing between CH and gold is rather slow with a characteristic time $\tau$ longer than hydrodynamic timescales. Inside the capsule, the mixing between CH and DT fuel is much faster for densities of order of 20\udens\, and temperatures of 1\,keV. Mixing times are of the order of tenths of ns, which means that mixing must be considered during implosions. Finally, at the high  temperatures and the very low densities in the hohlraum, the mixing between CH and the gas filling the capsule is very fast, with characteristic times of order of tens of picosecond. 

Table \ref{bin}  gives the diffusion coefficients computed considering the empirical average component approach denoted \De\, for different mixtures and those obtained by using Eq.\,\eqref{eq:effective} denoted \Dh. For the three mixtures [CH]Au, [CH][DT] and [CH]He, the diffusion coefficients   are computed with PIJ, either by considering the empirical binary mixture for \De, or, by doing the explicit multicomponent mixture with PIJ and applying Eq.~\eqref{eq:effective} for \Dh. One can see that these two estimations are always of the same order of magnitude (less than a factor 4 of difference). However, when looking closer, the differences are much larger in the kinetic regime (low density and/or high temperature) reaching factors between 2 and 4. Interestingly, the two evaluations are much closer in the strongly-coupled regime (high density and/or low temperature), where the faster convergence towards the hydrodynamic limit for coupled systems can be invoked \cite{HANS06}. Consequently, while it is tempting to simplify the mixture description by considering an average effective material, this should be done carefully depending on the level of accuracy required. For order of magnitude estimates,  this reduction is valid. On the other hand, it is not advisable if accurate results are needed, especially in the kinetic regime.

%---------------------------------------------------------------------------------------------------
\subsection{Separated reactants experiments}
%---------------------------------------------------------------------------------------------------
Dedicated experiments have been designed to study microscopic mixing in ICF experiments. Among them, the measurement of DT fusion reaction rates with separated reactants \cite{MURP16,ZYLS18}  have shown the dominant role of the microscopic mixing  over turbulent mixing. The interpretation of these experiments rests on an evaluation of deuterium diffusion both in the CD layer and in the HT gas.  In Ref.\,\cite{ZYLS18}, deuterium must first diffuse through  the CD in the shell at low temperature (1.5~keV) and, next, mix with HT inside the capsule at much higher temperature (4~keV). 

The DC layer  in the shell is a binary mixture, but deuterium in the H-T gas of the capsule is a three-components mixture (D-H-T). Zylstra uses the empirical mixture of H and T to form an average element ([HT]: Z=1, $\overline{A}$=2) and applies Kagan's kinetic formulation  \cite{KAGA14} to the effective binary mixture (D-[HT]). Note that,  this effective mixture is one-component  since deuterium and [HT] have the same atomic and mass numbers. 

When used with Zylstra's prescriptions with the Coulomb logarithm set to 6, for collisions with C only, and for relaxation coefficients set to 1, the PIJ model exactly reproduces the diffusion constants. Predicting with PIJ  the Coulomb logarithm, bounded by 1.65,  and accounting for collisions between all species we get  3 times higher diffusivity for deuterium in CD with $\De=\Dh=15\,\dmu$.  In the hotter HT mixture, deuterium's diffusion  given by the empirical binary mixture  $\De=1080\,\dmu$ is very close to the hydrodynamical prescription  $\Dh=1100\,\dmu$, both being a little smaller than Zylstra's estimation $\Dz=1590\,\dmu$.
\begin{table}[!t]
\caption{Deuterium diffusivity in \dmu=10 cm$^2$/s in the DC mixture and in the DHT mixture. $\De$ is the evaluation  in the empirical effective binary mixture and $\Dh$ the hydrodynamic effective binary mixture.}
 \begin{center}
\begin{tabular}{|c|c|c|c|c|c|c|c|}
\hline
Mixture	&$x$	&	$T$	& $\rho$	& \Dz \cite{ZYLS18}	&\De		& \Dh	\\
		&\%	&	eV	& \udens	&	\dmu			&	\dmu		& \dmu		\\
\hline \hline
D C		&50	& 1500	& 1		&	5.1			& 15			& 15				 \\
%		&	&		&		&				&			&			\\
%\hline
D H T	&50	& 4000	& 0.5		&	1590			& 1080		& 1100		 \\
%		&	&		&		&				&			&			\\

\hline
\end{tabular}
\end{center}
\label{fig:zylstra}
\end{table}

%---------------------------------------------------------------------------------------------------
\section{Conclusion}
%---------------------------------------------------------------------------------------------------
We studied  three-component plasma mixtures  in the warm dense regime (100-400 eV). We used results from orbital-free molecular dynamics simulations to get realistic structural and transport data. We set a procedure to compute the ionizations for each species  establishing an iso-electronic equilibrium between species. The ion charges are then used in a multi-components HNC procedure which accurately reproduces the structure of the mixture and in a global formulation of the transport coefficients (PIJ) that successfully predict transport coefficients such as self-diffusion, mutual diffusion and viscosity. The gathering of similar species into an effective binary mixture is also tested. 

We finally apply our global model to situations encountered in recent ICF experiments. The PIJ prediction of diffusivity gives credit to the important role it may play in the mixing of CH ablator with DT fuel as well as with H/He gas filling the hohlraum. The time scales of diffusion are comparable to those derived with Kagan's formulation \cite{KAGA14}.

 %---------------------------------------------------------------------------------------------------
\section{Acknowledgements}
%---------------------------------------------------------------------------------------------------
Work supported under the auspices of Science Campaigns 4 and 1 and the Advanced 
Technology Computing Campaign (ATCC) by the US Department of Energy through the Los
Alamos National Laboratory. Los Alamos National Laboratory is operated by Triad National
Security, LLC, for the National Nuclear Security Administration of U.S. Department of
Energy (Contract No. 89233218NCA000001).

This work has been done under the NNSA/DAM collaborative agreement P184 on Basic Science.

%---------------------------------------------------------------------------------------------------
\section{References}
%---------------------------------------------------------------------------------------------------
%\bibliographystyle{apsrev}
%\bibliography{/Users/jeanclerouin/Documents/biblio/biblio,/Users/jeanclerouin/Documents/biblio/BIBLIOS/biblioPA,/Users/jeanclerouin/Documents/biblio/BIBLIOS/SSA_Bibliographie} 

\begin{thebibliography}{44}
\bibitem[{Com()}]{Comment_Gam}
\emph{\bibinfo{title}{{The definition of the coupling parameter in a
  multispecies system is exactly the same as the one used in astrophysics,
  except that partial ionizations are considered here with $Q_i \leq Z_i$}}}.
\end{thebibliography}

 %---------------------------------------------------------------------------------------------------
 %---------------------------------------------------------------------------------------------------
 \appendix
 %---------------------------------------------------------------------------------------------------
  %---------------------------------------------------------------------------------------------------
\section{Multicomponent mutual diffusion in the Darken approximation}
 %---------------------------------------------------------------------------------------------------
 As shown by the previous study \cite{WHIT19}, the Darken approximation is valid to within a few percent in most  cases and furnishes a convenient framework to understand multicomponent diffusion. Let us recall that this approximation neglects cross terms in the corresponding correlation functions. The agreement with a direct calculation is generally very good but must be checked by numerical simulations. The N-components  formulation of the Darken approximation for mutual diffusion is \cite{WHIT19}
\begin{eqnarray}
   \label{Darken1}
D_{ij}&=&{\frac {D_iD_j}{D_{mix}}} \\   
 {\frac{1} {D_{mix}}}&=&\sum_i^{N}{\frac {x_i}{D_i}} .
   \label{Darken2}
\end{eqnarray} 
In this formula, the self-diffusion coefficients $D_i$ must be evaluated in the mixture.  They are noted $D_j$ for a pure element, $D_j^{(2)}$ for the self diffusion of element $j$ in a binary mixture, $D_j^{(3)}$  in a ternary mixture etc.

 Consider the mutual diffusion between two components 1 and 2 in a binary mixture, quoted $D_{12}^{(2)}$.
The Darken formula is
 \begin{eqnarray}
{\frac{1}{D_{mix}}}&=&{\frac{x_1}{D_1^{(2)}}} +{\frac{x_2}{D_2^{(2)}}} \\ \nonumber
&=&{\frac{x_1D_2^{(2)}+x_2 D_1^{(2)}}{D_1^{(2)}D_2^{(2)}}} ,
\end{eqnarray}
which gives the usual well-known Darken formula for  binary mixtures
 \begin{eqnarray}
\darken_{12}^{(2)} &=&x_1D_2^{(2)} +x_2D_1^{(2)}.
\end{eqnarray}
For 3 components we can rewrite Eq.\,(\ref{Darken1}) and (\ref{Darken2})
 \begin{eqnarray}
D_{12}^{(3)}&=&x_1D_2^{(3)}+x_2 D_1^{(3)}+x_3{\frac {D_1^{(3)}D_2^{(3)}}{D_3^{(3)}}} \\
\label{d123}	&=& \darken_{12}^{(3)} +x_3{\frac {D_1^{(3)}D_2^{(3)}}{D_3^{(3)}}} 
\end{eqnarray}
Note that $\darken_{12}^{(2)}$  is different from  $\darken_{12}^{(3)}$ since the self diffusion coefficients are respectively taken in the binary and in the ternary mixture. Omitting the superscript in Eq.~(\ref{d123}), we end up with a relation 
 \begin{equation}
D_{12}= \darken_{12} +x_3{\frac {D_1D_2}{D_3}} ,
\label{app:darken}
\end{equation} 
with all quantities taken in the  actual mixture. Eq.~(\ref{app:darken}) shows how the mutual diffusion between two components (in the whole mixture)   is modified by the adjunction of a third one. 
At high concentration in the heavy element ($x_1,x_2 \rightarrow 0$ and $x_3 \rightarrow 1$), one get a mutual diffusion $D_{12}$ bigger than $D_1$  if $D_3$ is smaller than $D_2$, which is always the case if the third element is a HZE.

%---------------------------------------------------------------------------------------------------
\section{Ionizations and couplings}
 %---------------------------------------------------------------------------------------------------
Mixture, total density, Wigner-Seitz radius $a$, ionic radii $a_1$, $a_2$, $a_3$, charges $Q_1$, $Q_2$, $Q_3$ and coupling parameters  $\Gamma_1$, $\Gamma_2$, $\Gamma_3$  under the  \iso\, prescription for a HXAg mixtures (0.4:0.4:0.2) with X=D, He, Li, C, Na, K, Cr, Cu and Rb at 200 (Table~\ref{table2}) and 400~eV (Table~\ref{table3}). 

\ISO\, prescription for a HCAg mixture ($x$:$x$:$1-2x$) versus $x$ at 200~eV (Table~\ref{table4}).

\begin{table*}[!h]
\caption{\ISO~  predictions for the global density in \udens, the Wigner-Seitz radius, the ionic radii (in atomic units), the charges and couplings of each species for various mixtures (0.4:0.4:0.2) at 200~eV and a constant pressure of 720\,Mbar.}
\begin{center}
\begin{tabular}{|c|c|c||c|c|c||c|c|c||c|c|c|}
\hline
Mix	& $\rho_\text{Tot}$   & $ a$   &  a$_\text{H}$    &  a$_\text{X}$    &  a$_\text{Ag}$   &  Q$_\text{H}$   &  Q$_\text{X}$    &  Q$_\text{Ag}$   &  $\Gamma_\text{HH}$    &  $\Gamma_\text{XX}$    & $\Gamma_\text{AgAg}$	\\
\hline\hline
{\bf HDAg} 	& 16.08 &  1.56 &  0.97 &  0.97 &  2.49 &  0.95 &  0.95 & 16.28 &  0.128 &  0.128 & 14.50\\
{\bf HHeAg} 	& 15.52 &  1.56 &  0.96 &  1.19 &  2.47 &  0.95 &  1.83 & 16.27&  0.128 &  0.382 & 14.56 \\
{\bf HLiAg} 	& 15.36 &  1.63 &  0.96 &  1.35 &  2.46 &  0.95 &  2.64 & 16.26&  0.129 &  0.707 & 14.60\\
{\bf HCAg} 	& 14.45 &  1.70 &  0.95 &  1.62 &  2.44 &  0.95 &  4.74 & 16.24&  0.130 &  1.885 & 14.69\\
{\bf HNaAg} 	& 14.46 &  1.79 &  0.94 &  1.86 &  2.42 &  0.95 &  7.39 & 16.22&  0.131 &  3.987 & 14.78\\
{\bf HKAg} 	& 15.07 &  1.88 &  0.93 &  2.07 &  2.41 &  0.95 & 10.38 & 16.21&  0.132 &  7.075 & 14.86\\
{\bf HCrAg} 	& 16.11 &  1.92 &  0.93 &  2.16 &  2.40 &  0.95 & 11.80 & 16.20&  0.132 &  8.779 & 14.89\\
{\bf HCuAg} 	& 16.97 &  1.95 &  0.93 &  2.22 &  2.39&  0.95 & 13.00 & 16.20&  0.132 & 10.33 & 14.92\\
{\bf HRbAg} 	& 18.90 &  2.00 &  0.93 &  2.31 &  2.39 &  0.95 & 14.59 & 16.20&  0.132 & 12.55 & 14.95\\
\hline
\end{tabular}
\end{center}
\label{table2}
\caption{Same as Table~\ref{table2} at 400~eV and a constant pressure of 1520\,Mbar.}
\begin{center}
\begin{tabular}{|c|c|c||c|c|c||c|c|c||c|c|c|}
\hline
Mix	& $\rho_\text{Tot}$   & $ a$   &  a$_\text{H}$    &  a$_\text{X}$    &  a$_\text{Ag}$   &  Q$_\text{H}$   &  Q$_\text{X}$    &  Q$_\text{Ag}$   &  $\Gamma_\text{HH}$    &  $\Gamma_\text{XX}$    & $\Gamma_\text{AgAg}$	\\
\hline\hline
{\bf HDAg}	& 13.83 &  1.64 &  0.93 &  0.93 &  2.66 &  0.98 &  0.98 & 23.17 &  0.070 &  0.070 & 13.72   \\
{\bf HHeAg}	& 13.50 &  1.67 &  0.92 &  1.16 &  2.65 &  0.98 &  1.92 & 23.15 &  0.070 &  0.216 & 13.75   \\
{\bf HLiAg}	& 13.45 &  1.70 &  0.92 &  1.31 &  2.65 &  0.98 &  2.83 & 23.14 &  0.070 &  0.414 & 13.77   \\
{\bf HCAg}	& 12.71 &  1.78 &  0.92 &  1.62 &  2.63 &  0.98 &  5.36 & 23.11 &  0.071 &  1.21 & 13.81   \\
{\bf HNaAg}	& 12.56 &  1.88 &  0.91 &  1.91 &  2.61 &  0.98 &  8.98 & 23.07 &  0.071 &  2.87 & 13.86   \\
{\bf HKAg}		& 12.76 &  1.99 &  0.90 &  2.17 &  2.60 &  0.98 & 13.50 & 23.04 &  0.072 &  5.71 & 13.91   \\
{\bf HCrAg}	& 13.45 &  2.04 &  0.90 &  2.28 &  2.59 &  0.98 & 15.75 & 23.02 &  0.072 &  7.40 & 13.93   \\
{\bf HCuAg}	& 14.01 &  2.08 &  0.90 &  2.37 &  2.58 &  0.98 & 17.70 & 23.02 &  0.072 &  9.00 & 13.95   \\
{\bf HRbAg}	& 15.38 &  2.14 &  0.90 &  2.47 &  2.58 &  0.98 & 20.40 & 23.00 &  0.072 & 11.36& 13.97   \\
\hline
\end{tabular}
\end{center}
\label{table3}
\caption{Same as Table~\ref{table2} for an equimolar mixture of H and C  with a varying proportion of silver at 200~eV and a constant pressure of 720\,Mbar}
\begin{center}
\begin{tabular}{|c|c|c||c|c|c||c|c|c||c|c|c|}
\hline
\%Ag	& $\rho_\text{Tot}$   & $ a$   &  a$_\text{H}$    &  a$_\text{C}$    &  a$_\text{Ag}$   &  Q$_\text{H}$   &  Q$_\text{C}$    &  Q$_\text{Ag}$   &  $\Gamma_\text{HH}$    &  $\Gamma_\text{CC}$    & $\Gamma_\text{AgAg}$	\\
\hline\hline
0.00	&  5.95 &  1.43 &  0.99 &  1.70 &  - &  0.95 &  4.77 & - &  0.13 &  1.83 & -   \\
0.05 	&  9.01 &  1.51 &  0.97 &  1.67 &  2.51 &  0.95 &  4.76 & 16.3 &  0.13 &  1.85 & 14.4   \\
0.10 	& 11.29 &  1.58 &  0.96 &  1.65 &  2.48 &  0.95 &  4.75 & 16.3 &  0.13 &  1.86 & 14.5   \\
0.20	 & 14.44 &  1.70 &  0.95 &  1.62 &  2.44 &  0.95 &  4.74 & 16.2 &  0.13 &  1.88 & 14.7   \\
0.25	& 15.58 &  1.76 &  0.94 &  1.61 &  2.43 &  0.95 &  4.73 & 16.2 &  0.13 &  1.89 & 14.8   \\
0.50	& 19.10 &  2.00 &  0.93 &  1.58 &  2.39 &  0.95 &  4.72 & 16.2 &  0.13 &  1.92 & 14.9   \\
0.75	& 20.93 &  2.19 &  0.92 &  1.57 &  2.37 &  0.95 &  4.72 & 16.2 &  0.13 &  1.93 & 15.0   \\
\hline
\end{tabular}
\end{center}
\label{table4}
\end{table*}

%---------------------------------------------------------------------------------------------------
  %---------------------------------------------------------------------------------------------------
\section{Hydrodynamical reduction}
\label{app:hydro}
We give the explicit formulation of Eq.\:(\ref{eq:effective}) for a ternary mixture turned in a 2+1 binary mixture and a 4-component mixture turned into a 2+2 binary mixture. It is easy to check that for for a binary mixture $\Dh=D_\text{12}$.
 %---------------------------------------------------------------------------------------------------
 \subsection{3-component mixture}
 %---------------------------------------------------------------------------------------------------
In a 3-component mixture (e.g. H-C-Au=1-2-3), we gather C and H into an effective [CH] element to compute the penetration of Au into CH. The mixture [CH]-Au defines $E_1=1, 2$ and $E_2=3$. If $\{x_i\}$ are the number fractions we have $y_i=\rho_i/\rho=x_i*A_i/\Amoy$, where $\Amoy=\sum{x_i A_i}$ and $\alpha_1=\alpha_2=0.5$ and $\alpha_3=1$. Given the mutual diffusion coefficients $D_{ij}$ computed with PIJ in the 3-component mixture, the effective diffusion between [CH] and Au is given by
\begin{equation}
\Dh=(\alpha_1 y_3+\alpha_3 y_1)D_{13}+(\alpha_3 y_2+\alpha_2 y_3) D_{23}.
\label{eq:3eff}
\end{equation}
The coefficient $D_{12}$ does not appear due to sum rules.

 %---------------------------------------------------------------------------------------------------
 \subsection{4-component mixture}
 %---------------------------------------------------------------------------------------------------
A D-T-H-C (1-2-3-4) mixture is reduced to a binary [DT]-[CH]. $D_{ij}$ mutual diffusion coefficients  are computed with PIJ in the 4-component mixture
 \begin{eqnarray}
\Dh&=&(\alpha_1 y_3+\alpha_3 y_1)D_{13}+(\alpha_3 y_2+\alpha_2 y_3) D_{23}   \nonumber \\
	&+& (\alpha_4 y_1+\alpha_1 y_4)D_{14}+(\alpha_4 y_2+\alpha_2 y_4) D_{24}. \nonumber \\
\label{eq:4eff}
\end{eqnarray}

 %---------------------------------------------------------------------------------------------------
 \subsection{Glossary}
 %---------------------------------------------------------------------------------------------------
\begin{itemize} [label=--]
\item BIM: binary ionic mixtures;
\item DFT; density functional theory;
\item DNS: direct numerical simulation;
\item FPL: Fokker-Plank Landau;
\item HNC: hyper-netted chain;
\item HZE: high Z element (W, Ag, Au,...);
\item ICF: inertial confinement fusion;
\item LZE: low Z element (H, He, ..);
\item MZE: medium Z element  (C,Al ...);
\item MCHNC: multi-components hyper-netted chain;
\item NS: Navier-Stokes;
\item OCP: one component plasma;
\item OFMD: orbital free molecular dynamics;
\item OZ: Ornstein-Zernicke relation;
\item PDF: pair distribution function;
\item PIJ: pseudo-atom in Jellium;
\item TIM: ternary ionic mixtures;
\item VACF: velocity autocorrelation function.
\end{itemize}

\end{document}